\documentclass[nofootinbib,superscriptaddress,eqsecnum,tightenlines,10pt,longbibliography,notitlepage]{revtex4-1}
\pdfoutput=1
\usepackage{hyperref}
\usepackage{graphicx}
\usepackage{amsmath,amssymb,amsfonts,amsthm,stmaryrd,mathtools,bbold,mathtools}
\usepackage{yfonts}
\usepackage{multirow,bigdelim}
\usepackage{dsfont}
\usepackage{color}
\usepackage{graphicx,color,epstopdf,epsfig,psfrag}
\usepackage[dvipsnames]{xcolor}
\usepackage{ mathrsfs }

\hypersetup{
    colorlinks=true,       
    linkcolor=MidnightBlue,          
    citecolor=BrickRed,        
    filecolor=BrickRed,      
    urlcolor=BrickRed           
}
 
\makeatletter
\newcounter{subsubsubsection}[subsubsection]
\renewcommand\thesubsubsubsection{\thesubsubsection .\@alph\c@subsubsubsection}
\newcommand\subsubsubsection{\@startsection{subsubsubsection}{4}{\z@}%
                                     {-3.25ex\@plus -1ex \@minus -.2ex}%
                                     {1.5ex \@plus .2ex}%
                                     {\centering\normalfont\small\textit}}
\newcommand*\l@subsubsubsection{\@dottedtocline{3}{10.0em}{4.1em}}
\newcommand*{\subsubsubsectionmark}[1]{}
\makeatother

\def\be{\begin{equation}}
\def\ee{\end{equation}}
\def\ba{\begin{eqnarray}}
\def\ea{\end{eqnarray}}
\def\bas{\begin{subequations}\begin{eqnarray}}
\def\eas{\end{eqnarray}\end{subequations}}

\def\tr{\text{tr}}

\def\SU{\text{SU}}

\def\nn{\nonumber}
\def\q{\qquad}

\def\wG{\widetilde{G}}
\def\wH{\widetilde{H}}
\def\mG{\mathcal{G}}
\def\mD{\mathcal{D}}
\def\mR{\mathcal{R}}

\def\uRoman#1{\uRomannumeral{\the\value{#1}}}

\def\ne{{n^{\rm e}}}

\def\nr{{n^{\rm r}}}

\renewcommand{\S}{{\mathbb S}}
\renewcommand{\tilde}{\widetilde}

\def\D{\mathfrak{D}}

\begin{document}

\title{\LARGE On entanglement entropy in non-Abelian lattice gauge theory and 3D quantum gravity}

\author{Clement Delcamp}\email{cdelcampATperimeterinstituteDOTca}
\affiliation{Perimeter Institute for Theoretical Physics,\\ 31 Caroline Street North, Waterloo, Ontario, Canada N2L 2Y5}
\affiliation{Department of Physics $\&$ Astronomy and Guelph-Waterloo Physics Institute \\  University of Waterloo, Waterloo, Ontario N2L 3G1, Canada}
\author{Bianca Dittrich}\email{bdittrichATperimeterinstituteDOTca}
\affiliation{Perimeter Institute for Theoretical Physics,\\ 31 Caroline Street North, Waterloo, Ontario, Canada N2L 2Y5}
\author{Aldo Riello}\email{arielloATperimeterinstituteDOTca}
\affiliation{Perimeter Institute for Theoretical Physics,\\ 31 Caroline Street North, Waterloo, Ontario, Canada N2L 2Y5}

\begin{abstract}

Entanglement entropy is a valuable tool for characterizing the correlation structure of quantum field theories. When applied to gauge theories, subtleties arise which prevent the factorization of the Hilbert space underlying the notion of entanglement entropy. Borrowing techniques from extended topological field theories, we introduce a new definition of entanglement entropy for both Abelian and non--Abelian gauge theories.  Being based on the notion of excitations, it  provides a completely relational way of defining regions. Therefore, it naturally applies to background independent theories, e.g. gravity, by circumventing the difficulty of specifying the position of the entangling surface. We relate our construction to earlier proposals and argue that it brings these closer to each other. In particular, it yields the non--Abelian analogue of the `magnetic centre choice', as obtained through an extended--Hilbert--space method, but applied to the recently introduced fusion basis for 3D lattice gauge theories. We point out that the different definitions of entanglement entropy can be related to a choice of (squeezed) vacuum state.

\end{abstract}
\maketitle

\section{Introduction}

Entanglement entropy has become an important tool for characterizing the correlation structure of quantum field theories \cite{Calabrese:2004eu,Amico,Casini:2009sr}, in particular with regard to correlations in space. In the latter case  one presupposes that field degrees of freedom can be localized. 

Gauge theories, however, feature a form of non-locality that prevents the strict localization of the so--called physical, as opposed to gauge-variant, degrees of freedom.  For instance, in Yang Mills theories, including electromagnetism, the presence of  Gau\ss~constraints implies that one can compute  the total electric charge contained in a region  solely in terms of the electric flux across the region's boundary: no information about the bulk fields is needed.

Quantum mechanically, this non-locality is reflected in the fact that the Hilbert space of gauge-invariant states does not factorize into the tensor product of Hilbert spaces associated to a spacetime region $A$ and its complement $B$. More precisely  the algebra of gauge invariant observables does not factorize into the product of two commuting subalgebras each containing only operators supported in either $A$ or $B$.
Consequently, the definition of the entanglement entropy between one region and its complement requires further discussion, especially in the light of the privileged role gauge theories play in nature.

%
In quantum gravity this problem appears  even more cogent, see also the recent discussions \cite{Giddings2015}.
This happens not so much because a complete quantum theory of gravity is yet to be defined and agreed upon (in three dimensions one can actually argue for the opposite), but rather because of the very defining property of gravity: background independence. Indeed, because of background independence, which implies diffeomorphism invariance, the localization of regions and their separation into distinguished subsystems, when performed from {\it within} the theory itself, is already a thorny subject.
To address the definition of entanglement entropy in a background-independent fashion, we advance a proposal which we believe sheds light also onto some of the issues encountered already within the standard gauge-theoretical framework.



To start with let us discuss the extant proposals for lattice gauge theories.  In this context, essentially two approaches have been proposed for how to define entanglement entropy between two regions. Our work will further relate these two approaches and provide a new point of view.

The first approach \cite{Donnelly2008,Buivi,Donnelly2011,Donnelly2014}, in particular put forward by Donnelly, is based on the embedding of the Hilbert space of gauge invariant states---which displays the non-local features discussed above---into an extended Hilbert space where gauge-invariance violations are allowed at the interface. This extended Hilbert space does factorize, allowing to define the entanglement entropy of a gauge invariant state as the entropy of its embedding. 
 This proposal works for Abelian as well as non-Abelian gauge theories.  

 We will show here that there are different possible extension procedures.  In fact, the procedure chosen in \cite{Donnelly2008,Buivi,Donnelly2011,Donnelly2014}  relates to a choice of vacuum state describing the strong coupling limit of lattice gauge theory.  In this work we will propose an alternative extension procedure---this time related to the weak coupling limit---leading to an alternative definition of the entanglement entropy.  

The ambiguity in the notion of entanglement entropy for lattice gauge theories has been pointed out by Casini, Huerta and Rosabal (CHR) in \cite{Casini2013} and discussed in a framework that focuses on the algebra of gauge invariant observables. This was fully developed only for Abelian gauge theories in \cite{Casini2013}.\footnote{The ``electric'', as opposed to ``magnetic'', prescription has been generalized to the non-Abelian case by \cite{Soni2015}, and we will comment on this in section \ref{subsec_observ}. In this work we will also generalize the  ``magnetic'' prescription, and show that it actually is not purely magnetic in the non--Abelian case.}
In order to define a notion of entanglement entropy between two regions $A$ and $B$ in the algebraic framework one starts with the algebra ${\cal O}$ of (gauge invariant) observables and seeks to split this algebra into two mutually commuting subsets, associated to region $A$ and $B$ respectively. The difficulty is however that due to the non--locality of gauge invariant observables, there is a set of observables which can be associated to both regions. This might be both due to ({\it e.g.} string like) observables crossing the boundary and due to observables being localized on the boundary. The ambiguity in the definition of entanglement entropy comes from a choice of a subset of such observables, which is {\it removed} from the observable algebra. The reduced algebra ${\cal O}_{\text{red}}$ features then (generically) a centre ${\cal Z}$, basically consisting of the observables conjugated to the removed ones. The resulting observable algebra should be such that the observables in the centre ${\cal Z}$ can be associated to the boundary, whereas the remaining observables ${\cal O}_{\text{red}}\backslash {\cal Z}$ can be associated to either region $A$ or $B$. 

The gauge-invariant Hilbert space, which carries an (irreducible) representation of the algebra ${\cal O}$, will now feature superselection sectors with respect to ${\cal O}_{\text{red}}$: the common eigenspaces of the centre ${\cal Z}$ are left invariant by ${\cal O}_{\text{red}}$. At the same time each eigenspace admits a factorization into Hilbert spaces associated to region ${A}$ and ${ B}$.  The definition of entanglement entropy can be extended to such cases with superselection sectors and thus provides the notion of entanglement entropy in this algebraic framework.

Furthermore, the superselection sectors  label states with a definite choice of boundary conditions, that is with definite values for the observables in the centre ${\cal Z}$. The latter set is sometimes referred to as boundary observables, a viewpoint that is also reflected in the discussion of \cite{DonnFreid}  on (classical) gauge systems with boundary.  As we will argue in section \ref{sec_ent}, in the non--Abelian case one can extend the set of gauge-invariant boundary observables by including non--gauge invariant frame information  attached to the boundary. Indeed, this step is necessary to completely match the algebraic definition of entanglement entropy to the one based on an Hilbert space extension: in the non-Abelian case the extension includes also the addition of the frame information.

The two most natural choices for boundary conditions in the Abelian case consist in either fixing the electric fluxes on links transversal to the boundary, or the magnetic fluxes (Wilson loops) along the boundary. Correspondingly, CHR speak of an electric or magnetic choice of centre. This can also be matched to different ways of cutting the lattice into two parts, either across or along the links.  We will show here that the generalization of the magnetic centre choice to the non--Abelian case does also add an electric component to the centre, measuring the total electric flux through the boundary.\footnote{This observable is vanishing in the Abelian case due to the Gau\ss~constraints. In non--Abelian gauge theories one might get `effective' electric charges on larger scales without having an electric charge at the lattice scale. Such charges were called `Cheshire charges' in \cite{Bais1991}.}
Moreover, in the magnetic-centre case, no extended Hilbert space procedure was known. Our proposal fills this gap in the Abelian case and also provides the generalization to the non-Abelian one.

What might result surprising is the striking discrepancy in the behaviour of the entanglement entropy with respect to different choices of boundary conditions. This happens even when the entropies are calculated for the same state and on the same geometry. The authors of \cite{Casini2013}  argue that all these ambiguities disappear in the continuum limit, once only well--defined quantities such as relative entropies are considered. Let us, however, point out that---in a continuum limit appropriate to describe states of a topological field theory with defect excitations---the different procedures even yield finite or divergent results respectively. 

This brings us to the philosophy pursued here: we will focus our attention not so much on the lattice and how it is separated into two distinguished regions, but on the excitation content of the theory under consideration. Importantly the very notion of `excitation' depends on a choice of vacuum state. We will encounter two choices of totally squeezed vacua, which have definite values either for the electric fluxes (here assumed to all vanish)  or the magnetic fluxes (also assumed to vanish). In the loop quantum gravity literature, these vacua are known as Ashtekar--Lewandowski (AL) \cite{Ashtekar1986,Ashtekar1991,Ashtekar1993} and BF vacua \cite{DGflux,DGfluxC,DGfluxQ}, respectively; they are related to the strong and weak coupling regime of lattice gauge theory, respectively.

Using such vacua, one can interpret  states on a finite lattice as states with a finite number of excitations in a continuum theory.  This is done by basically putting all degrees of freedom, finer than the lattice scale, into the chosen vacuum \cite{DittrichSteinhaus13}.  In this picture, the choice of boundary conditions specifying a notion of entanglement entropy should be adjusted to the choice of vacuum.  For states which have almost everywhere vanishing electrical field we should choose electric boundary conditions. Considering states which have almost everywhere vanishing magnetic flux---or in the gravitational language have almost everywhere vanishing curvature---we should choose instead the magnetic centre. These choices are designed to give finite results in the continuum limit, whose result is by construction already fully captured at the level of a finite (fine enough) lattice.

 Furthermore, our definition of excitations is rooted in the analysis of the properties that regions with boundaries manifest under gluing. In fact, we will also argue that the process of cutting a system in two and the related definition of an extended Hilbert space procedure should be understood  as dual to gluing. This brings into play techniques of extended topological field theory, {\it i.e.} topological field theory for manifolds with boundaries. 
 
 In particular, the choice of the BF vacuum will naturally lead us to consider a different basis for the Hilbert space, the `fusion basis'.  
Importantly, thanks to a precise definition of  its excitations, we can operationally specify regions by their excitation content.
In this way we overcome the problems arising from defining a region independently of its content (something that would be in stark contrast with background independence), and we do so in a way that makes such a definition independent of the chosen regularizing lattice, thus directly avoiding the need of specifying the way we split it. 

Indeed, our proposal comes to full fruition in $(2+1)$-dimensional gravity. This can be formulated as a gauge theory and moreover does not feature propagating degrees of freedom. Indeed it can be formulated as a topological theory, known as BF theory (from which the BF vacua are derived). This theory can be coupled to point particles which carry exactly the type of excitations considered in this paper.  Therefore, regions will be specified by the point particles they contain.  In this sense, our notion of entanglement theory characterizes the correlations between the excitations the regions contain. 

We will focus our discussion  to $(2+1)$-dimensional lattice gauge theories, and restrict for simplicity to finite gauge groups. A strategy to generalize the fusion basis, employed here, to $(3+1)$ dimensions, can be found in \cite{DD16}. The generalization to the Lie-group case requires the definition of the so--called Drinfel'd double for such groups, which can be found in \cite{Koornwinder1996,Koornwinder1998}.


The paper is organized as follows.  In section \ref{sec_splitting}, we present the gluing of spatial manifolds with boundary. We then define the splitting procedure dual to this gluing as well as the notion of extended Hilbert space. In section \ref{sec_ent}, we present the extended Hilbert space method of computing entanglement entropy and emphasize the relation with the observable algebra approach. In section \ref{sec_fusion}, we introduce the set-up for lattice gauge theory and perform a change of point of view by focusing on the excitation content of the theory instead of the underlying lattice. This leads to the construction of the fusion basis, which will be key to our definition of extended Hilbert space. A new notion of entanglement entropy for lattice gauge theories is introduced in section \ref{sec_ent_lat} where several explicit calculations are presented. Finally, in section \ref{sec_ent_gravity}, we discuss the implications of this new definition for the case of 3D quantum gravity.


\section{Surface gluings and splittings, and extended Hilbert spaces \label{sec_splitting}}

To define the entanglement entropy of a subregion, one first needs to specify how to associate the theory's degrees of freedom to it. In gauge theories, having to consider non-local gauge-invariant degrees of freedom, this leads to ambiguities. 

The gauge-invariance condition manifest itself in terms of constraints, {\it i.e.} quantum versions of the elliptic equations a state must satisfy to represent valid initial data. In lattice gauge theories, this leads to Gau\ss~constraints defined at the lattice nodes. The Gau\ss~constraint at the node $n$ involves all the links adjacent to it, links which carry gauge co-variant degrees of freedoms expressed in terms of parallel transports (holonomies) along open paths. As a consequence, a gauge-invariant wave function necessarily correlates the degrees of freedom across the links. 

In turn, this prevents the splitting of the Hilbert space of gauge-invariant functions ${\cal H}$  into a tensor product ${\cal H}_A \otimes {\cal H}_B$, with the two factors associated to two complementary regions.  The so--called extended Hilbert space procedure \cite{Donnelly2011,Donnelly2014} avoids this task, by considering an extended Hilbert space ${\cal H}_{\rm ext}$, in which the Gau\ss~ constraints are relaxed along the boundary interface between the two regions. More precisely, this interface is defined to be transversal to the links of the lattice, and a two--valent node is introduced on each link cut by it. The Gau\ss~constraints are then relaxed for these two-valent nodes only. This defines an extended Hilbert space, which does factorize ${\cal H}_{\rm ext}=  {\cal H}_A \otimes {\cal H}_B$ in a straightforward manner.

In this work we will show that the extended Hilbert space procedure can be generalized using a different set--up and also different sets of constraints. Furthermore this generalized procedure is deeply connected to the theory of extended topological field theories. There, indeed, one considers topological field theories on manifolds with boundaries\footnote{ Here we are working in a Hamiltonian framework, therefore `boundaries' have to be understood as codimension 2 surface, which---in a covariant context---are usually called `corners'.} together with a procedure for gluing them to one-another. Splitting a manifold into two components thus arises as an inverse procedure. 

In the following we will describe the main idea and start with the notion of gluing states, defined on spatial manifolds with boundary. Dual to this gluing one can define a splitting procedure and the notion of extended Hilbert spaces.

~\\
\subsection{Gluing}
To be concrete, we consider a theory where the (gauge co--variant) degrees of freedom are associated to the links of a graph $\Gamma$, as is the case in lattice gauge theories, where one has group elements $g_l$ associated to the links $l$ of $\Gamma$. 

Let $\Gamma_A$ and $\Gamma_B$ be two graphs embedded into the hypersurfaces, $\Sigma_A$ and $\Sigma_B$ respectively . We assume both $\Sigma_A$ and $\Sigma_B$ have boundaries and that the embedded graphs end at the boundaries by one or several open edges. The (gauge--covariant) wave functions defined on $\Gamma_A$ and $\Gamma_B$ respectively live in the kinematical Hilbert spaces $\mathcal{H}^{\text{kin}}_{A}$ and $\mathcal{H}^{\text{kin}}_{B}$. Let then $\{ \mathscr{C} \}_A$ be a set of constraints  which we require to be quasi-local, {\it i.e.} local {\it e.g.} with respect to the graph's nodes and faces and their adjacent structures, and $\mathcal{H}_A$ the subspace of wave functions in the Hilbert space $\mathcal{H}_{A}^{\text{kin}}$ which satisfies these constraints.\footnote{Here we make the simplifying assumptions that the set of solutions to the constraints can inherit the inner product of $\mathcal{H}_{A}^{\text{kin}}$. This happens if zero is in the discrete spectrum of the constraints.  If this is not the case a new inner product needs to be constructed, see {\it e.g.} \cite{RAQ1,RAQ2, MCP1,MCP5}, and the following procedure needs to be amended accordingly.} Likewise for $\{ \mathscr{C} \}_B$. To every constraint $\mathscr{C}$, we assign a projector $\mathbb{P}_{\mathscr{C}}$ which projects onto the subspace of wave functions $\psi$ satisfying the constraint: $\mathscr{C}\psi=0$. 

As an example, we mentioned the Gau\ss~constraint above, which acts at the nodes of the graph imposing gauge invariance. We consider however only the Gau\ss~constraint acting at the internal nodes, that is everywhere {\it except at the (one-valent) nodes on the boundaries}\footnote{Imposing a Gau\ss~constraint for these one--valent nodes would mean to trivialize the dependence of the wave function on the group element associated to the link ending at the boundary.} of $\Sigma_A$ or $\Sigma_B$.

At the level of the surfaces, the gluing between $\Sigma_A$ and $\Sigma_B$ is obtained by identifying a portion of their boundaries. We denote the result of this operation $\Sigma_{A\cup B}$. At the level of the embedded graphs, it is analogously defined by connecting the links along which the gluing is performed.  We denote the result $\Gamma_{A\cup B}$. Here we assume that the links ending at the two boundaries match under the gluing procedure.\footnote{This can be ensured by introducing marked points on the boundaries where the links are allowed to end.} 
Let us now define the gluing operation {\it for two wave-functions} and denote it with a star $\star: \mathcal{H}_{A} \otimes \mathcal{H}_{B}\rightarrow   \mathcal{H}_{A\cup B} $, where the wave functions in $\mathcal{H}_{A\cup B}$ have to satisfy the set of constraints $\{{\mathscr{C}}\}_{A\cup B}$, which we specify presently.  The $\star$ gluing is defined in two steps.

Given two wave functions $\psi_A \in \mathcal{H}_A$ and $\phi_B \in \mathcal{H}_B$, consider first the (usual $\mathbb C$-)product of wave functions $\psi_A \cdot \phi_B$, defined on the glued graph $ \Gamma_{A  \cup B}$.  
In general, this product wave function will {\it not} satisfy all the constraints $\{\mathscr{C}\}_{A\cup B}$, which will include $\{ \mathscr{C}\}_A$ and $\{ \mathscr{C}\}_B$ but also further constraints that result from the presence of new internal nodes and faces in $\Gamma_{A \cup B}$.  Nevertheless, the set of wave functions of the form $\psi_A \cdot \phi_B$ will span the extended Hilbert space  $\mathcal{H}_{A} \otimes \mathcal{H}_{B}=:{\cal H}_{\text{ext}}$, and hence the subspace $\mathcal{H}_{A\cup B}$ can be identified with the set of subspace of wave functions which satisfy all the constraints $\{\mathscr{C}\}_{A\cup B}$. Denoting the corresponding projector by $\mathbb{P}_{A\cup B}$ we finally define the star product as
\be\label{starprod}
	\begin{array}{cccccl}
		\star:& \;{\cal H}_A \otimes {\cal H}_B& \xrightarrow{{\;\; \cdot \;\;}} & {\cal H}_{\text{ext}} & \xrightarrow{{\mathbb P}_{A\cup B}} &{\cal H}_{{A\cup B}} \\
				& (\psi_A , \phi_B) &  \xmapsto{\phantom{\;\; \cdot \;\;}} & \psi_A\cdot \phi_B& \xmapsto{\phantom{{\mathbb P}_{A\cup B}}} &  {\mathbb P}_{A\cup B} ( \psi_A \cdot 			\phi_{B})
	\end{array}
\ee
In the following we will denote for brevity, $C = A\cup B$.


\subsection{Splitting}

Splitting is the `inverse' operation of gluing. Given a surface $\Sigma_C$ and embedded graph $\Gamma_C$ we firstly have to introduce a boundary that splits $\Sigma_C$ into $\Sigma_A$ and $\Sigma_B$ and $\Gamma_C$ into $\Gamma_A$ and $\Gamma_B$ so that the gluing gives back the corresponding structures. (One might want to choose certain restrictions on which kinds of boundaries and graphs are allowed.) 

To define the splitting of a wave function in ${\cal H}_C$ we are looking for an isometric embedding map 
\ba
{\cal E}:  {\cal H}_C \rightarrow {\cal H}_{\text{ext}} \simeq \mathcal{H}_A \otimes \mathcal{H}_B \, ,
\ea
such that $\star \circ {\cal E} = \mathrm{id}$ on ${\cal H}_C$. Note that this latter condition does not specify ${\cal E}$ uniquely, but we can demand that ${\cal E}$ maps $ {\cal H}_C$ into ${\cal H}_C$ understood as a subspace of ${\cal H}_{\text{ext}}$. (Remember that ${\cal H}_C$ can be identified with the set of wave functions in ${\cal H}_{\text{ext}}$ satisfying all the constraints $\{\mathscr{C}\}_C$.)

The embedding ${\cal E}$, therefore, maps wave functions in ${\cal H}_C$, which does not allow a straightforward splitting, into an extended Hilbert space ${\cal H}_{\text{ext}} \simeq \mathcal{H}_A \otimes \mathcal{H}_B$, which comes with a natural tensor factorization associated to the splitting of $\Gamma_C$ into $\Gamma_A$ and $\Gamma_B$. Hence, to integrate out the degrees of freedom associated {\it e.g.} to $\Gamma_B$, one first uses the embedding, and then traces over $\mathcal{H}_B$.

\subsection{Flatness constraint}
We have already mentioned the Gau\ss~constraint of gauge theories.
However, for our proposal the introduction of another constraint will be relevant. This is the flatness constraint, which acts at the faces of the graph and demands that these have
a trivial holonomy,\footnote{That is $g_{l_N} \cdots g_{l_1}=e$ with $e$ denoting the group unit and the group elements following clockwise or anti--clockwise order along the boundary of the face.} {\it i.e.} a trivial magnetic flux through them.
Again, we will have closed faces, necessarily internal to $\Sigma$, and open faces as well, necessarily including boundary components. 
We will demand the flatness constraints to hold for closed faces only. 

At this point the reader might wonder why we are interested in the flatness constraints. Firstly, we can allow for curvature by introducing punctures, that is by removing disks from $\Sigma$ and thus introducing boundaries around which the flatness constraint does not need to hold. Introducing sufficiently many punctures we can regain all curvature degrees of freedom. Therefore, our procedure is not over-restrictive. On the other hand, the introduction of flatness constraints allows to achieve a certain independence from the lattice by putting the focus rather on the punctures themselves, which provide the support for the excitations.  
Secondly, with regard to the gluing and extension process, the flatness constraints allow us to trade all Gau\ss~constraint violations, appearing in the extended Hilbert space as defined in \cite{Donnelly2011,Donnelly2014}, for one flatness  and Gau\ss~constraint violation. The reason is that with the flatness constraints holding almost everywhere (except at the punctures) we can change the graph and its embedding without changing the physical content of the wave functions. Hence, in this way we can change the number of links crossing the boundary to just one link. In other words, the many local reference frames defined by the cut links are replaced by a global reference frame together with demanding a locally flat connection near the boundary. We will later see that this allows as to define the (generalization of the)  `magnetic centre' choice \cite{Casini2013} in terms of an extended Hilbert space procedure.

\subsection{Gluing of cylinders}
Let us mention a further interesting application of the gluing procedure, which is crucial for the definition of the fusion basis and the notion of excitations.
Consider a two--punctured sphere, denoted ${\mathbb{S}}_2$. It is topologically equivalent to a (hollow) cylinder. Let us denote by ${\cal H}_{\mathbb{S}_2}$ the Hilbert space of wave functions on a graph embedded in ${\mathbb{S}}_2$, satisfying the flatness and Gau\ss~constraints for closed faces and inner nodes, respectively. Notice, that two copies of $\mathbb S_2$ can be glued together giving another $\mathbb S_2$. Therefore, the state-gluing operation $\star$ defines a multiplication map in $\mathbb S_2$:
\ba
\star: {\cal H}_{\mathbb{S}_2} \otimes {\cal H}_{\mathbb{S}_2} \rightarrow {\cal H}_{\mathbb{S}_2} \; .
\ea 
This equips ${\cal H}_{\mathbb{S}_2}$ with the structure of an algebra. 
More precisely, it turns out \cite{DDR1} that the $\star$ multiplication provides a representation of the multiplication of the Drinfel'd double algebra ${\cal D}({\cal G})$ of the gauge group ${\cal G}$ \cite{drinfel1988}. The (irreducible) representations of this Drinfel'd double algebra will play a central role in our proposal,  in particular in the explicit construction of the fusion basis ({\it cf.} section \ref{subsub_fusion}).

\section{Entanglement entropy \label{sec_ent}}

We reviewed the issues arising when attempting the splitting of a Hilbert space ${\cal H}_C$ of wave functions satisfying a set of constraints $\{{\mathscr C}\}_C$  into a tensor product.  Such a splitting can be performed by  embedding the states in an extended Hilbert space ${\cal H}_{\text{ext}} \simeq {\cal H}_A \otimes {\cal H}_B$ for which some constraints are relaxed.  This is described by an embedding map $\mathcal{E}: {\cal H}_C \rightarrow {\cal H}_{\text{ext}}$.

With a choice of embedding map  at hand, we can define a notion of entanglement entropy for states in ${\cal H}_C$.  To do so, we first use the map ${\cal E}$ to embed a given state $\psi\in{\cal H}_C$ into  ${\cal H}_{\text{ext}}\simeq {\cal H}_A \otimes {\cal H}_B$, hence we define the reduced density matrix
\be
\D^A_\psi = \tr_B \big({\cal E}(\psi)\overline{{\cal E}(\psi)} \big) \, ,
\label{eq_densitymatrix}
\ee
from which the entanglement entropy can be readily evaluated
\be\label{ent1}
	S_A(\psi)\,:=\,S_A( {\cal E}(\psi)) \,=\, - \tr_A ( \D^A_\psi \ln \D^A_\psi) \;  .
\ee
Notice that both $\D^A$ and $S_A$ implicitly depend on $\cal E$.

In \cite{Donnelly2008,Donnelly2011,Donnelly2014}, a definition of entanglement entropy was proposed for both Abelian and non-Abelian gauge theories by Donnelly. His procedure was the type we just described---often referred to as the `extended Hilbert space' method---and made implicit use of a specific embedding map. In \cite{Casini2013}, CHR pointed out that (at least in the Abelian case) Donnelly's procedure agrees with their `electric centre' prescription, but it was just one among other choices.

Here, we want to emphasize that, by choosing embedding maps different from Donnelly's, the extended Hilbert space construction can be generalized and is therefore not unique.  In particular, the alternative procedure proposed here does reproduce CHR's `magnetic centre' prescription. This at least for Abelian gauge theories, since we will see that the non-Abelian case necessarily includes also an electric component.
Hence, in so doing, we provide a tighter connection between CHR's algebraic constructions and the extended Hilbert space procedure. Moreover, by explicitly providing an extended Hilbert space procedure matching the `magnetic' centre prescription, we correct claims about its impossibility which have appeared in the literature \cite{DonnellyGiddings2016}.

In the rest of this section, we will describe in detail the contributions to the entanglement entropy, as defined by the extended Hilbert space procedure, along the lines of Donnelly \cite{Donnelly2014}. While his analysis was based on a specific embedding procedure (corresponding to a choice of spin network basis for the Hilbert spaces involved), we will instead allow for generic embeddings and associated choices of basis. With this toolbox at hand, we will relate the extended Hilbert space procedure to CHR's observable-algebra-based definition \cite{Casini2013}. It is left to the forthcoming sections, the task of introducing the details of the fusion basis for $(2+1)$ dimensional lattice gauge theories \cite{DDR1}, needed to parallel the magnetic centre choice, and the study of the corresponding embedding procedure and entanglement entropy.

\subsection{Entanglement entropy from extended Hilbert spaces}

Both the spin network basis and the fusion basis are indexed by representation labels, which we here will denote generically by $\rho$. Notice, however, that these are representations for different algebraic structures, namely for the group ${\cal G}$ and its Drinfel'd double ${\cal D}({\cal G})$, respectively ({\it cf.} section \ref{subsub_fusion}).  As usual for basis-state labels, the $\rho$'s encode the eigenvalues of a maximal set of commuting observables on the Hilbert space ${\cal H}_C$. This hints already at the connection to CHR's observable-algebra-based procedure as well as to more general choices of maximal sets of commuting observables.

We split the representation labels into three sets: $\{\rho_A\}$, associated to region $A$; $\{\rho_B\}$, associated to region $B$; and $\{\rho_\partial\}$ associated to the boundary $\partial A=\partial B$. 
In the case of the fusion basis we will just have one $\rho_\partial$ associated to the boundary.  Also, in case we allow torsion excitation at the punctures---that is violations of gauge invariance there---we have representation space indices $I$ associated to these punctures. These can be associated either to region $A$  or $B$ and we will therefore subsume them into the set of representation indices $\{\rho_A\}$ and $\{\rho_B\}$ respectively. 

In the extended Hilbert space ${\cal H}_{\text{ext}}={\cal H}_A \otimes {\cal H}_B$ we will have a basis that includes a doubling of the $\{\rho_\partial\}$ labels to $\{\rho_{\partial A}\}$ and $\{\rho_{\partial B}\}$. Furthermore, for each of these label sets, we have associated sets of representation-space labels\footnote{ By this we mean the following. Call $V_\rho$ the vector space supporting the representation $\rho$, then $I$ labels the elements of a basis of $V_\rho$.} $\{I_{\partial A}\}$ and $\{I_{\partial B}\}$. 

Denote $| \rho_A,\rho_B, \rho_\partial\rangle$ and
$|\rho_A, \rho_{\partial A}, I_{\partial A} \rangle \otimes |\rho_B, \rho_{\partial B}, I_{\partial B} \rangle$ elements of an orthonormal basis of ${\cal H}_C$ and ${\cal H}_A\otimes {\cal H}_B$, respectively. The embedding map $\cal E$ is then  given by
\ba
{\cal E}\, | \rho_A,\rho_B, \rho_\partial\rangle =  \frac{1 }{\prod_\partial \sqrt{\text{dim} (\rho_{\partial })}} \sum_{I_{\partial }}      \, |\rho_A, \rho_{\partial }, I_{\partial } \rangle \otimes |\rho_B, \rho_{\partial }, I_{\partial } \rangle \; .
\ea
(The product is over the boundary elements. In the case of the fusion basis we will have only one boundary element and index $\rho_\partial$. In the case of the spin network basis any edge cut by the boundary is a boundary element.) 

Given a state
\ba
|\psi \rangle&=& \psi( \{\rho\} ) \,  | \rho_A,\rho_B, \rho_\partial\rangle  \; \in \; {\cal H}_C\,,
\ea
the corresponding density matrix $\D^A_\psi$, defined in \eqref{eq_densitymatrix}, has the following structure:
\ba
\D^A_\psi &=& \bigoplus_{ \rho_\partial}  P(  \rho_{\partial})  \left[   \frac{  
|\rho_\partial ,I_\partial \rangle  \langle \rho_\partial, I_\partial |
 }
 { \prod_ \partial \text{dim}(\partial \rho)}  \,\, \otimes \D^A_\psi( \rho_\partial) \right] .
\ea
We see that $\D^A$ is block diagonal, with each block labeled by a boundary-representation vector $|\rho_{\partial A}= \rho_\partial, I_{\partial A}=I_\partial \rangle$, and weighted by the probability distribution
\ba
P(  \rho_{\partial}) = \sum_{\rho_A, \rho_B} \psi( \{\rho\}) \overline{  \psi( \{\rho\})   }.
\ea
This distribution is constant over the $I_\partial$ as a consequence of gauge invariance.  
On the other hand, the density matrix associated to each block (independent of $I_\partial$)  is
\ba
\D^A_\psi( \rho_\partial) &=&  \sum_{\rho_A,\tilde \rho_A, \rho_B} \frac{1}{P(\rho_\partial)} \psi(\rho_A, \rho_B, \rho_\partial) \overline{ \psi( \tilde\rho_A , \rho_B, \rho_\partial  )  } \,\,   | \rho_A \rangle \langle \tilde \rho_A | .
\ea
Now, given this decomposition, one finds that the entanglement entropy \eqref{ent1} has three contributions \cite{Donnelly2011}
\ba\label{ent2}
S_A\,=\, H( P(\rho_\partial)) \,+\, \sum_\partial \langle \ln \dim \rho_\partial \rangle \,+\, \langle S_A( \D^A(\rho_\partial)) \rangle , 
\ea
where $\langle {\scriptstyle \bullet} \rangle$ stands for the expectation value with respect to the classical probability distribution $P(\rho_\partial)$, and $H( P(\rho_\partial))$ for its Shannon entropy,
\ba
H( P(\rho_\partial)) \,=\, - \sum_{\rho_\partial} P(\rho_\partial )\, \ln P(\rho_\partial).
\ea

\subsection{Relation to observable-algebra-based entanglement entropy \label{subsec_observ}}

We now comment on the relation between this approach and the definition of entanglement entropy via the splitting of the observable algebra. CHR's original proposal \cite{Casini2013} concerned only Abelian gauge theories. We will comment below on the non-Abelain generalizations.  

Given the algebra of observables, ${\cal O}$, associated to the gauge-invariant Hilbert space ${\cal H}_C$, one chooses a commuting subalgebra of observable, ${\cal Z}$, associated to the boundary $\delta A=\delta B$.  This commuting subset ${\cal Z}$ of observables serves as centre of a new, reduced, observable algebra ${\cal O}_{\text{red}}$, which is obtained by removing all the observables, which do not commute with the designated centre ${\cal Z}$. 

The choice of $\cal Z$ must be done in such a way that ${\cal O}_{\text{red}}$ admits a splitting ${\cal O}_{\text{red}}= {\cal O}_A \cup {\cal O}_B$ into two mutually commuting subalgebras, which can be associated to the regions $A$ and $B$, respectively. These subalgebras clearly have a non-vanishing intersection given by the centre, ${\cal O}_A \cap {\cal O}_B={\cal Z}$.

Now, ${\cal H}_C$  (usually) provides an irreducible representation of the observable algebra ${\cal O}$. By removing observables from ${\cal O}$, one finds that the reduced algebra ${\cal O}_{\text{red}}$ features superselection sectors on ${\cal H}_C$. These superselection sectors are precisely labeled by the eigenvalues $\{\lambda\}_{\cal Z}$ for the observables in ${\cal Z}$.  This is because the original Hilbert space has by construction the structure
\ba\label{ssstructure1}
{\cal H}_C\,=\, \bigoplus_{\{\lambda\}} {\cal H}^{\{\lambda\}}_C \,=\,  \bigoplus_{\{\lambda\}} {\cal H}^{\{\lambda\}}_A \otimes  {\cal H}^{\{\lambda\}}_B \; ,
\ea
where each superselection sector ${\cal H}^{\{\lambda\}}_C$ can be readily factorized into ${\cal H}^{\{\lambda\}}_A \otimes  {\cal H}^{\{\lambda\}}_B$. Clearly, ${\cal H}^{\{\lambda\}}_{A}$ (${\cal H}^{\{\lambda\}}_{B}$) carries a representation of the algebra ${\cal O}_A$ (${\cal O}_B$, respectively), with elements in ${\cal Z}$ acting trivially as multiples of the identity operator. 

The observables in ${\cal Z}$ can be also understood as  boundary conditions, characterizing each of the superselection sectors.  This interpretation physically explains the `classical' behaviour of these observables noticed by CHR. See also the discussion in \cite{Gomes2016}.

At the beginning of this section, we introduced the basis $|\rho_A,\rho_B, \rho_\partial\rangle$ for ${\cal H}_C$. This basis immediately suggests one to choose  the centre ${\cal Z}$ to be generated by the projectors ${\mathbb{P}}_{\rho_\partial}$ onto the subspaces spanned by the $|\rho_A,\rho_B, \rho_\partial\rangle$ with varying $\rho_A,\rho_B$ but fixed $\rho_\partial$.  
This is equivalent to requiring the eigenvalues $\{\lambda\}_{\cal Z}$ to be directly determined by the labels $\rho_\partial \equiv \{\rho_\partial \}_{\cal Z}$. Henceforth, with this choice in mind, we will replace the superindex $\{\lambda\}$ by $\rho_\partial$.

The definition of entanglement entropy via the specification of a centre by CHR \cite{Casini2013} did originally concern only the Abelian case. It can also be generalized to the non--Abelian case, albeit in two different manners. One choice corresponds to staying withing the algebraic framework based on gauge-invariant observables alone \cite{Soni2015}.  In this case one forms density matrices with a superselection structure as given by \eqref{ssstructure1}
\ba
\D \,=\, \bigoplus_{ \rho_\partial}  \,  P( \rho_\partial)  \, \D(\rho_\partial) \; .
\ea
Factorizing each sector ${\cal H}^{\rho_\partial}$ into ${\cal H}^{\rho_\partial}_A \otimes {\cal H}^{\rho_\partial}_B$ one can compute the entanglement entropy $S_A({\D}^A(\rho_\partial))$ for each sector separately. The entanglement entropy for the entire system is then defined as
\ba\label{ent3.13}
S_A:= H( P(  \rho_\partial)) + \langle S_A(\D(\rho_\partial)) \rangle,
\ea
where again $\langle {\scriptsize \bullet} \rangle$ stands for the averaging with respect to $P(\rho_\partial)$. 

We see that this result does not completely reproduce the extended Hilbert space procedure (\ref{ent2}), as in (\ref{ent3.13}) we do not have the term $\langle  \ln \dim \rho_\partial \rangle$ appearing (this term trivially vanishes for the Abelian case).
The source of the discrepancy is the following.
In the extended Hilbert space procedure, the density matrices resulting from the embedding map ${\cal E}$, have also a superselection structure. But this superselection structure is more refined:  additional subsectors appear which are related to the internal indices of the representation  $I_{\partial}:=I_{\partial A}=I_{\partial B}$. Consequently, the density matrices resulting form the embedding procedure are {\it effectively} characterized by the following probability distribution
\ba\label{SSS1}
P(\rho_\partial,I_\partial) =   \frac{P(\rho_\partial) }{ \prod_\partial \dim \rho_\partial}   ,
\ea
and block density matrices
\ba
\D^A(\rho_\partial, I_\partial)\,=\,\frac{ \D^A( \rho_\partial) }{{\prod_\partial }\dim \rho_\partial   }.
\ea
The entanglement entropy for density matrices with such a superselection structure is given by 
 \ba\label{SSS2}
 S_A &=& H( P(\rho_\partial,I_\partial))  \,+\, \langle  S_A( \D^A(\rho_\partial, I_\partial)) \rangle_{P(\rho_\partial,I_\partial)} \nn\\
 &=& H( P(\rho_\partial)) \,+\, \sum_\partial \langle \ln \dim \rho_\partial \rangle_{P(\rho_\partial)} \,+\, \langle S_A( \D^A(\rho_\partial)) \rangle_{P(\rho_\partial)}
 \ea
where here $\langle {\scriptstyle \bullet} \rangle_{P(\rho_\partial,I_\partial)}$ denotes averaging with respect to $P(\rho_\partial,I_\partial)$ and $\langle {\scriptstyle \bullet} \rangle_{P(\rho_\partial)}$ averaging with respect to $P(\rho_\partial)$. This definition reproduces the splitting into three terms as in (\ref{ent2}).  

This second choice of superselection structure, which includes the magnetic indices $I_\partial$,  is not tied to the initial gauge invariant observable algebra, as the magnetic indices only arise after cutting the manifold.  One can, however, argue that cutting the manifold one introduces a boundary, and that the magnetic indices should be part of the boundary data together with the  $\rho_\partial$, characterizing (sectors of) wave functions defined on manifolds with boundary.  In other words, one could argue  that the splitting of a system into subsystem requires the introduction of additional information about the  reference frames at the boundary, as encoded in the magnetic indices, which is needed to perform a consistent gluing.

 Thus the first two contributions to the entanglement entropy in \eqref{SSS2} are resulting from the superselection structure and are thus due to the classical probability distribution \eqref{SSS1}. Indeed it was conjectured by CHR and proven by \cite{Soni2015, Verstraete2015}  that only the third contribtion in \eqref{SSS2} gives the so--called distillable entropy, which is defined to be the maximum number of Bell pairs that can be extracted by a so--called entanglement distillation. The latter process involves a choice of (local) operator algebra, which in \cite{Soni2015,Verstraete2015} is based on the reduced operator algebra ${\cal O}_{\text{red}}$.   Thus the notion of distillable entropy also depends on the choice of reduced operator algebra or alternatively boundary conditions.

From a physical standpoint, what all this discussion is  reminding us is that the concept of entropy is coarse--graining, {\it i.e.} observer, dependent. By varying the amount of information we know, or conversely, we would like to know about a system, we calculate different entropies. This can be summarized in the statement, that entropy is an epistemological quantity. And in sophisticated enough situations, also the entanglement entropy is such.

As mentioned, the extended Hilbert space procedure was first proposed using spin network functions \cite{Buivi,Donnelly2008, Donnelly2011, Donnelly2014}. Here the representation labels $\rho_\partial$ characterize the eigenvalues of electric flux operators associated to the links that are cut by the boundary (for non--Abelian gauge theories one can take Casimir operators formed from the electric fluxes associated to each such link, see \cite{Soni2015}).  Thus, in the corresponding algebraic definition the centre is formed by these electric operators.
 
 In this paper, we focus on the extended Hilbert space procedure for the fusion basis, where $\rho_\partial$ characterizes a so-called  closed-ribbon operator along the boundary between the two regions $A$ and $B$. In the case of an Abelian theory (and considering only gauge-invariant wave functions), this ribbon operator reduces to a Wilson loop. Thus, in this case, the centre is given by a `magnetic' operator. 
 For non--Abelian gauge theories, however, the closed-ribbon operator measures also an electric excitation, related to the total flux of the electric field flowing out of the enclosed region.  Note that  this can be non--trivial even for completely gauge invariant wave functions.  Thus, for non-Abelian theories, the magnetic centre gets naturally enlarged by a further `electric' operator.
 
Finally, we now turn to the introduction of the fusion basis.


\section{Fusion basis for lattice gauge theories \label{sec_fusion}}

In this section, we review the construction of the fusion basis for a $(2+1)$--dimensional lattice gauge system. For a more extensive treatment we refer the reader to \cite{DDR1}. For simplicity, we will assume that the gauge group is a finite group $\mathcal G$. We will also fix the topology of the underlying two--dimensional hypersurface to be spherical ($\mathbb S$) with possibly $p$ punctures present, {\it i.e.} $\Sigma\simeq\mathbb{S}_p$. 

\subsection{Hilbert space ${\cal H}_\Gamma$}
 
Let  $\Gamma$ be a graph embedded in $\S$. To start with, assume that the graph has no open links, {\it i.e.} no links ending at one--valent nodes.  The graph gauge-connection is defined by associating a group element to every (oriented) link of the graph corresponding to the holonomy along the link. The Hilbert space $\mathcal{H}_{\Gamma}$ is spanned by  the functionals $\psi : \mathcal{G}^L \rightarrow \mathbb{C}$ on the space of holonomies, where $L$ denotes the number of links in $\Gamma$. The Hilbert space $\mathcal{H}_{\Gamma}$ is equipped with an inner product defined as 
\ba\label{innprod1}
	\langle \psi_1, \psi_2\rangle  &=& \frac{1}{|{\cal G}|^{L}} \sum_{\{g_l\}} \overline{\psi_1( \{g_l\}}) \, \psi_2(\{g_l\}) .
\ea
Gauge transformations are parametrized by $\{u_n\}_n \in \mathcal{G}^N$, where $n$ denotes a node of $\Gamma$ and $N$  the number of such nodes. A gauge transformation acts on a holonomy configuration $\{g_l\} \in \mathcal{G}^L$ as
\ba
	\left( \{u_n\}_n  \triangleright \psi \right) (\{g_l\}) \,=\, \psi (\{ u^{-1}_{t(l)} \, g_l  \, u_{s(l)} \}),
\ea
where $s(l)$ and $t(l)$  denote the source and the target nodes of the link $l$, respectively. Gauge invariant functions are functions invariant under this gauge action. This defines a subspace  ${\cal H}_\Gamma^{\cal G} \subset \mathcal{H}_{\Gamma}$ of gauge invariant functions in the Hilbert space ${\cal H}_\Gamma$. ${\cal H}_\Gamma^{\cal G}$ inherits the inner product \eqref{innprod1} from ${\cal H}_\Gamma$. 
The gauge-invariance condition is encoded in the following Gau\ss~constraints (or projectors), associated to the nodes $n$:
\ba\label{Gauss1}
	\left({\mathbb P}^n_{\rm gauge} \psi \right) ( \{g_l\}) \,=\, \sum_u     \psi( \{u^{-1} g_{t}\}_t,\{g_{s} u \} , \{g_{l'}\}_{l'}  )   \,\, \stackrel{!}{=} \psi(\{ g_l\}) ,
\ea
where $s$  and $t$ index the links for which $n$ is a source and target node, respectively, while $l'$ indexes the remaining links.

\subsection{Basic operators}

Considering the configuration space to be the space of group holonomies, we have available two kinds of operators, namely Wilson loop operators and translation operators. 

Wilson loop or (closed) holonomy operators $W^f_\gamma$ act as multiplication operators on states $\psi(\{g\})$. Given a function $f:\mathcal{G}\rightarrow \mathbb{C}$ and a path $\gamma$  which coincides with some oriented and connected path along the links of $\Gamma$, the action of $W^f_\gamma$ is given by
\be
	(W^f_\gamma \psi) (\{g\}) \,=\, f(h_\gamma) \, \psi (\{ g\})  ,
\ee
where $h_\gamma = g_{l_n} \cdots g_{l_1}$ for $\gamma=l_n \circ \cdots \circ l_1$.  Note that for $W^f_\gamma$ to commute with the Gau\ss~constraints, $\gamma$ has to be a closed path and $f$ a class function. 

Translation operators $T_k[H]$ act by finite translations and therefore correspond to an exponentiated version of flux operators. We can define both a left and a right action for such operators. We choose to work with left translation operators whose explicit action is given by
\be
	(T_k[H]\psi)(g_1, \cdots, g_L) = \psi(g_1,\cdots,H^{-1}g_k,\cdots,g_L)  .
\ee
The action of $T_k[H]$ typically induces violations of the Gau{\ss} constraint at the target node of $l_k$, {\it {\it i.e.}} $t(l_k)$. 
However, the node at which the Gau\ss~constraint violation occurs can be moved at will. For this, one can parallel transport the to-be-translated argument $g_k$  from its  target node $t(l_k)$ to some other node $n$ along a path\footnote{This path should not include the link $l_k$ itself.} $\gamma$, apply the translation in the frame of $n$, and then transport the resulting holonomy back. We denote these operators $T_{k,\gamma}[H]$ and their action reads
\be
	(T_{k,\gamma}[H]\psi)(g_1, \cdots, g_L) = \psi(g_1,\cdots,h_\gamma^{-1}H^{-1}h_\gamma g_k,\cdots,g_L) ,
\ee
where $h_\gamma$ was defined above as the holonomy along the path $\gamma$.
The violation of the Gau{\ss} constraint induced by the translation now appears at the node $n$. This property will be important in the forthcoming construction where we will combine these basic operators in order to obtain so-called `ribbon operators'.

\subsection{Shift of viewpoint}

Let $\Gamma$ be a graph embedded into $\mathbb{S}$. 
$\Gamma$ being planar, we can unambiguously identify its plaquettes or faces. The shift of point of view we propose relies on the assumption that $\Gamma$ carries excitations located at the faces. As we will explain presently, these excitations have to be understood with respect to a given vacuum. 

First, we consider curvature excitations since they are naturally carried by the faces of $\Gamma$. Indeed, they are characterized by the amount of curvature carried by every face, defined as the trace of the holonomy surrounding the face. In the electromagnetic case these are precisely the magnetic fluxes.

\begin{figure}[h]
	\includegraphics[scale =0.8]{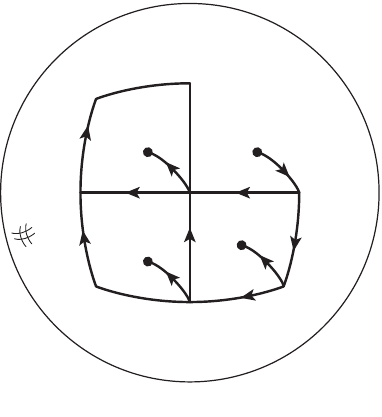} \q 
	\includegraphics[scale =0.8]{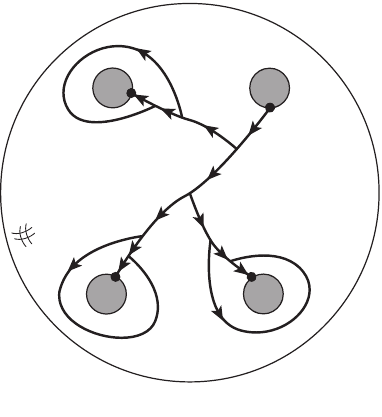} \,
	\includegraphics[scale =0.8]{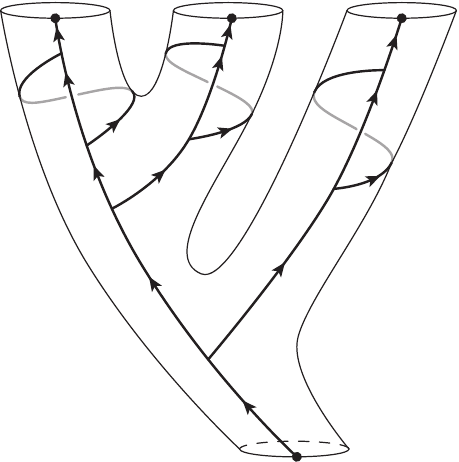} 
	\caption{The left panel represents a lattice of three plaquettes embedded on a two-sphere by closing it with an outer plaquette. For every plaquette, including the outer one, an open edge going from the lattice to a marked point can carry torsion degrees of freedom. The middle panel represents an equivalent description where the plaquettes are replaced by punctures. The solid lines now represent a minimal graph embedded on the corresponding punctured sphere. The right panel finally proposes another graphical representation where the topology of the punctured sphere is deformed to obtain pairs of pants. This final representation is the preferential one for the construction of the fusion basis. }
\label{fig_lattice}
\end{figure}

Then, we consider torsion excitations, that is violations of the Gau\ss~constraints \eqref{Gauss1}. In the electromagnetic case these excitations correspond to the presence of non-vanishing electric charges. 
Being associated with a Gau\ss~constraint violation, these excitations are {\it a priori} located at the nodes of $\Gamma$, and not at its faces as we desired. 
To obviate this problem, we introduce extra links and nodes. More precisely, we introduce one new link and one new node for each face  (in the context of combinatorial quantization of Chern Simons theory, this structure is called a `cilium').  For a given face this new link starts at some node on its boundary and ends at a new one--valent node placed in its interior. Henceforth, we refer to these nodes as `end nodes' $\{n^{\rm e}\}$, and to all others nodes as `internal nodes' $\{n^{\rm i}\}$. The valency of an internal node is strictly bigger than one. In the same spirit, we call the links adjacent to the end nodes `open links'. Figure \ref{fig_lattice} depicts such a construction in the case of a lattice with square faces. 

The result of this construction is an extended graph $\Gamma'$ which leads to a new Hilbert space ${\cal H}_{\Gamma'}$ equipped with an inner product of the same form
as the previous one, see  \eqref{innprod1}.

As it will become clear later on, allowing for torsion excitations is a necessity in the case of non-Abelian gauge theories, even in the case we do not allow them at the lattice scale.  We restrict our focus on the subspace of ${\cal H}_{\Gamma'}$ constituted by wave functions which are gauge invariant at all internal nodes, but not at the end nodes. This defines a new Hilbert space, ${\cal H}_p$, where $p$ stands for the number of end-nodes in $\Gamma$, which by construction coincides with the number of its faces, too:
\be
	{\cal H}_p=\{\psi \in {\cal H}_{\Gamma'} \,|\, \big({\mathbb P}^{n^{\rm i}}_{\rm gauge}  \psi \big) = \psi , \,\, \forall  u \in {\cal G} , \,\,\forall    n^{\rm i} \in \Gamma'\} \; .
\ee
Note that the Hilbert space ${\cal H}_p$ is unitarily equivalent to the subspace of wave functions in ${\cal H}_{\Gamma}$ which are gauge invariant at all nodes where one does {\it not} attach an open link. In other words, we can map the torsion excitations from ${\cal H}_p$ to ${\cal H}_\Gamma$  by associating them with the nodes to which one attaches an open link. For the example of the lattice depicted in Figure \ref{fig_lattice} gauge invariance violations at almost all nodes can be taken into account in the Hilbert space ${\cal H}_p$. Furthermore one can also generalize the definition of ${\cal H}_p$, allowing more than one open link to end in a given face \cite{DDR1}. This allows to take into account all possible gauge invariance violations, starting from an arbitrary graph $\Gamma$.  

The change of point of view we adopt here can be made more explicit by placing a puncture in the middle of each face. More precisely, instead of thinking of a lattice embedded in $\S_2$ and allow for some excitations, we can directly imagine a graph embedded onto a punctured sphere. The punctures act as defects which are the only possible support for both curvature and torsion. In this case, we can map a lattice with $p$ faces to a graph embedded on a $p$-punctured sphere. The face holonomy becomes the holonomy surrounding the puncture while the open edges now go from a node of the graph to one-valent node sitting at the puncture. This correspondence is detailed in the next paragraph.

\subsection{ From graph to punctures, and the flat vacuum}

The notion of excitations is bound to a notion of vacuum, which  here is the state without any curvature and torsion excitations. This vacuum is a gauge invariant state peaked on flat connections, also known as BF vacuum.\footnote{For the sphere the first fundamental group is trivial, so local flatness implies global flatness.}

Now, imagine that one wishes to describe configurations of (continuum) connection fields that are everywhere flat, except at a pre--defined number of points. Including torsion excitations, we need to extend these points to so--called punctures, that is infinitesimal disks with a marked point on their boundaries. 
The connection degrees of freedom can now be encoded in a Hilbert space ${\cal H}_p$ as described above. Note however that the precise choice of graph does not matter. For instance due to the local flatness of the connection, we can deform links of the graph, as long as we are not crossing over a puncture.  

Moreover we can even allow for graph refinements, that is add links, so that we have additional faces, that do not contain a puncture. In this case we just need to make sure that all allowed wave functions prescribe vanishing curvature at each closed face ({\it i.e.} at each face with no associated puncture) and, similarly, that  they are gauge-invariant at all internal nodes.
In conclusion,  we are allowed to change of graph, as long as it is sufficiently fine to ($i$) capture the first fundamental group of the punctured sphere, and to ($ii$) allow at least one connected path between any pair of punctures.  See \cite{DDR1} for the precise transformation maps. 

Later, we will also introduce `ribbon' operators whose action on the states does not depend on the particular choice of underlying graph either. These are the operators operators which will be used to characterize the fusion basis. 

Let us emphasize that this is a useful viewpoint one can adopt which makes explicit the connection to topological field theory with defects. However, although convenient, it is not necessary, and one can also proceed by having the usual fixed lattice in mind.

\subsection{Holonomy basis for ${\cal H}_p$}

We now construct a holonomy basis of the Hilbert space ${\cal H}_p$.  
As the name suggests, this basis is designed to diagonalize holonomy operators. These operators are demanded to be based on paths which start and finish at the end nodes. To define a maximal set of such holonomy operators, we join the two following subsets: 
\begin{itemize}
\item[{\it i})]{\it G-holonomies}---%
First, we single out one end-node and call it the `root node', $n^{\rm r}$. We call the face enclosing this root node the `outer face'.
We then need to choose a set of paths from the root node to each of the other end nodes $\{n^{\rm e}\}$. For this we pick a (connected) spanning tree in $\Gamma'$ denoted ${\cal T}'$. Such a tree uniquely determines  a path $\mathcal{P}'_{n}$ from the root node to any other node $n$, and {\it a fortiori} also to the end nodes of $\Gamma'$. The set of $G$-holonomies $\{G_{n^{\rm e}}\}$ is defined as the oriented product of holonomies following the paths $\mathcal{P}'_{n^{\rm e}}$:
\be 
	G_\ne \equiv \prod^{\rightarrow}_{l \subset \mathcal{P}'_{n^{\rm e}}}g_l \; .
\ee
This set automatically fixes all holonomy between pairs of end nodes along paths supported on ${\cal T}'$. \\

\item[{\it ii})]{\it H-holonomies}---%
The second set is constituted of holonomies $\{H_{n^{\rm e}}\}$ based on closed paths $\{\mathcal{L}'_{n^{\rm e}}\}$, going anti--clockwise along the boundary of every face containing a puncture $n^{\rm e}$ (all the others being trivial, anyway) and starting at the end-node $n^{\rm e}$ associated to the face itself,
\be
	H_{\ne}\equiv  \prod^{\rightarrow}_{l \subset \mathcal{L}'_{n^{\rm e}}}g_l \; .
\ee
\end{itemize}

Note that in order to obtain a maximal set of holonomies, it is not necessary to include the one around the outer face as long as we include the holonomies around all the other faces.

\begin{figure}[h]
	\includegraphics[scale =1]{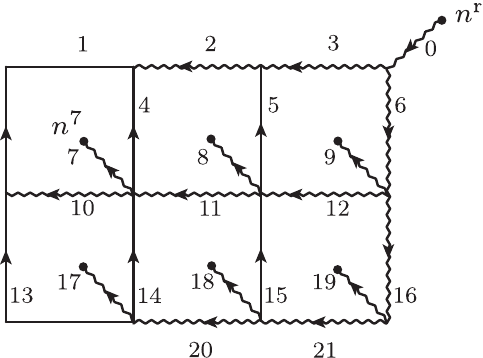} \q 
	\caption{Example of lattice with six plaquettes embedded on the two-sphere $\S_2$ by introducing an outer plaquette. The wiggly lines represent the connected spanning tree $\mathcal{T}'$. For each plaquette we associate a $G$-holonomy defined as the product of holonomies going from the corresponding end node to the root following $\mathcal{T}'$ and a $H$-holonomy defined by going anti-clockwise around the face starting and ending at the end-point. For the upper-left face we have for instance $G_7 = h_7h_{11}h_{12}h_6h_0$ and  $H_7 = h_7h_4^{-1}h_1h_{10}h_7^{-1}$. }
\label{fig_GHhol}
\end{figure}

Thus, a basis wave-function turns out to be labeled  by $(p-1)$ pairs $(G_\ne, H_\ne) \in {\cal G}^2$. Denote  it $\psi_{\{G_\ne,H_\ne\}}$. Figure \ref{fig_GHhol} depicts an example of such a construction. The wave functions can finally be written in a fully covariant form as a product over delta functions prescribing the $G$ and $H$-holonomies. For the sake of clarity, let us look at the minimal examples of a lattice with two faces. Replacing the faces by punctures, this corresponds to considering an embedded graph on the two-punctured sphere. Since the two-punctured sphere $\S_2$ is topologically equivalent to a cylinder, we have the following gaphical correspondence:
\be 
	\begin{array}{c}\includegraphics[scale =1]{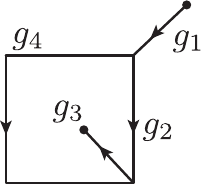}\end{array}\; \Leftrightarrow \;
	\begin{array}{c}\includegraphics[scale =1]{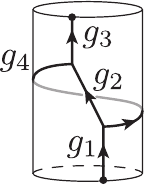}\end{array} \; .
\ee
with the marked point at the bottom puncture chosen as the root node.
Applying the previous prescriptions, the gauge covariant form for the holonomy basis states on the two--faces square lattice (or two--punctured sphere $\S_2$) is given by
\be\label{holb}
	\psi_{G,H}(g_1,\cdots,g_4)\,=\,  |\mathcal{G}|^{3/2} \delta(G,g_3g_2g_1)\delta(H,g_3g_4g^{-1}g_3^{-1}) \equiv
	\begin{array}{c}\includegraphics[scale =1]{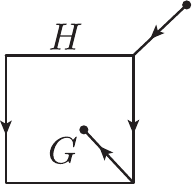}\end{array} = 
	\begin{array}{c}\includegraphics[scale =1]{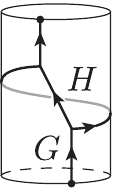}\end{array}    \,,
\ee
where we have chosen a particular normalization that will turn out to be convenient later on.
All delta-functions are Kronecker deltas, whose value is either zero or one.

\subsection{ Fusion basis and ribbon operators}

The holonomy basis $\{\psi_{\{G_{\ne},H_{\ne}\}}\}$ diagonalizes holonomy operators that are not gauge invariant at the end nodes and for this reason it is for now quite involved to specify a complete and independent subset of fully gauge invariant wave functions. Therefore, we first aim to find a (maximal) set of gauge-invariant operators and hence the basis which diagonalizes it. 

Starting from the holonomy basis, the previous remark suggests that we should include into the set of gauge invariant operators the conjugacy class of the holonomies $\{h_{n^{\rm e}}\}$. Let us for instance consider a lattice and two faces associated with the end nodes $n^{\rm e}_1$ and $n^{\rm e}_2$. We denote $h_{n^{\rm e}_2 \cup n^{\rm e}_1}$ the holonomy surrounding these two faces. If we have a non--Abelian group ${\cal G}$, knowing only the conjugacy classes $C_1$ and $C_2$ of the two holonomies $h_{n^{\rm e}_1}$ and $h_{n^{\rm e}_2}$, will in general  not determine the conjugacy class of $h_{n^{\rm e}_2 \cup n^{\rm e}_1}$ uniquely. 
Therefore the conjugacy class of the holonomy going around two faces generally encodes more information than the one provided by the conjugacy classes of the individual faces. It turns out that, knowing the individual conjugacy classes, the set of conjugacy classes one can obtain for the holonomy around the two faces is determined by so--called fusion rules.  

We propose to construct a basis which relies on the notion of fusion sketched above. For this reason, we refer to it as the fusion basis. The fusion basis diagonalizes a hierarchical set of (gauge invariant) operators, detecting the conjugacy classes of loop based holonomies. This hierarchical set is described by a so-called fusion tree. We choose it to be rooted and binary ({\it i.e.} with three--valent internal vertices) such that the end vertices of this tree are associated to the faces (or punctures) together with their corresponding end nodes, and the root of the tree is associated to the outer face with the  root node $\nr$. The combinatorial structure of the fusion tree determines which faces (or loops, or punctures) and in which order, are fused to form larger ones. Thus, the fusion tree determines for which hierarchical merging of loops one considers the associated closed holonomies. As an extra condition, we require the set of loops underlying the closed holonomies not to cross each other.

The hierarchical set of loops $\{\ell\}$ defined above prescribe gauge invariant functionals $\{f(g_{\ell})\}$ which detect the conjugacy classes and therefore capture the curvature (or magnetic) degrees of freedom. In particular, this defines Wilson loop operators $\{W^f_{\ell}\}$. However, we would also like to have operators that characterize the torsion (or electric) degrees of freedom. Indeed, even if we consider completely gauge-invariant functionals without torsion degrees of freedom for the original faces, we might have `emergent'  torsion degrees of freedom which arise when applying the fusion scheme described above. This is the one reason why torsion excitations might appear under coarse graining \cite{deWildPropitius,Livine2013,Charles2016, DGfluxC, DDR1}. This feature is again characteristic of non--Abelian groups and such effective torsion charges have been named Cheshire charges in \cite{deWildPropitius}. Conveniently, the torsion degrees of freedom can be captured with operators based on the same hierarchical set of loops, as the one used for the curvature degrees of freedom. The difference is that these operators include the action of translation operators. 

By putting together these two kinds of operators, we obtain the so--called ribbon operators introduced (in a slightly different form) by Kitaev \cite{Kitaev1997}. These operators measure both curvature and torsion excitations. We are now ready to review their construction.   

\subsubsection{Closed ribbon operators}

Let us now introduce the closed ribbon operators which are diagonalized by the fusion basis. We will see later that we can also define open ribbon operators. In order to describe the action of the closed ribbon operators, we will focus on a simple example, the general case as well as more details can be found  in \cite{DDR1}. 

\begin{figure}[h]
	\includegraphics[scale =1]{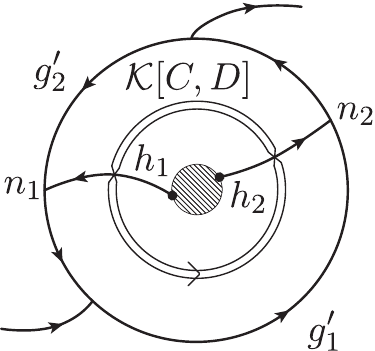} \q 
	\caption{Example of action of a closed ribbon operator. The ribbon operator, represented by the doubled represented line, acts on the links which it crosses. The dashed region accounts for the presence of any punctures and any graph. }
	\label{closedribbon}
\end{figure}

Consider the piece of graph displayed in Figure \ref{closedribbon}. There is a loop based holonomy given by $g_{\ell} = g'_2 g'_1$ where $g_2'$ is the composed holonomy which goes from $n_2$ to $n_1$ while $g_1'$ goes from $n_1$ to $n_2$. We are going to define the action of a directed closed ribbon operator along this loop $\ell$. By convention, the ribbon operator is drawn to the right (with respect to the orientation of the ribbon) of the associated loop $\ell$. For the purpose of describing the action of the ribbon operator we have to choose an initial node, which here will be $n_1$ {\it {\it i.e.}} the target node of the link carrying $h_1$. However, the action of the closed ribbon will eventually not depend on this choice.
 
Firstly, consider the action of the following (auxiliary) operator, parametrized by $(G,H) \in {\cal G}\times {\cal G}$:
\be\label{rib1}
	\left({\cal R}[G,H] \psi \right)(g'_1, g'_2, h_1,h_2,\cdots)
	=\delta(G,  g'_2g'_1) \, \psi(g'_1,g'_2,     (g'_2 g'_1)^{-1} H^{-1}  g'_2 g'_1 \, h_1 ,\,
	 (g'_2)^{-1} H^{-1} g'_2 \, h_2, \cdots) \;.
\ee
which is a combination of a holonomy operator---the wave function is multiplied by  $\delta(G,  g'_2g'_1)$---and a series of translation operators acting on the links crossed by the ribbon.  The translation parameter is always parallel transported along the loop $\ell$ to the node $n_1$. Both the holonomy action and the translational action {\it a priori} violate gauge-invariance at this node. Furthermore, as the ribbon contains a translational part, it might induce (or modify existing) curvature, at the face which includes the holonomy $h_1^{-1} g_2' h_2$. Indeed, this holonomy undergoes the shift
\be
\label{rib2}
	h_1^{-1} g_2'h_2 \,\,\rightarrow \,\,h_1^{-1} G^{-1} H G H^{-1}g_2' h_2 .
\ee
On the other hand, note that the holonomy $ h_2^{-1} g'_1 h_1$ stays invariant. 

Therefore the auxiliary operator ${\cal R}[G,H] $ might induce a change in curvature, via the action on the holonomy $h_1^{-1} g_2'h_2$, as well as violations of the gauge invariance at the node $n_1$. However, a closed ribbon operator is expected to only measure excitations, not to induce them. To obviate this problem, on the one hand we demand that $G$ commutes with $H$, that is we require $H$ to be in the stabilizer group $N_C$ of any representatives of the conjugacy class $C$ of $G$. This prevents any curvature modifications.
On the other hand, to deal with the gauge-invariance violation, we group-average the resulting wave function over the gauge action at the node $n_1$. This amounts to consider the averaging over the adjoint action of the group on the parameters $(G,H)$ of the ribbon. Putting everything together, one can show that the group averaged operator only depends on the conjugacy class $C$ of $G$ and a conjugacy class $D$ of $H \in N_C$: 
\ba
\label{rib3}
\sum_{h \in {\cal G}} {\cal R}[h G h^{-1}, h H h^{-1}]  
&=:&|N_D| \; {\cal K}[C,D]\,,
\ea
where the normalization factor $|{ N}_D|$ corresponds to the cardinality of the stabilizer group $N_D$ of $H \in N_C$. As expected the action of the operator ${\cal K}[C,D]$ does not depend any more on the choice of auxiliary node (which in this example was $n_1$). One can further show that the precise positioning of the ribbon with respect to the graph $\Gamma'$ is immaterial and the only information that matters about its position is topolgoical---{\it i.e.} it only matters how the ribbon winds around the locations of the excitations, that is the punctures. We have thus defined the closed ribbon operator $\cal K[C,D]$. 

It is already clear that the closed ribbon operator projects onto states peaked sharply on the conjugacy class $C$ for the loop based holonomy $g_{\ell}=g_2'g_1'$. We would also like to achieve a projector property with respect to the parameter $D$, prescribing the translational action of the ribbon operators on holonomies transversal to the loop holonomy. Usually, the diagonalization of a translation operator requires some sort of Fourier transform. 
Indeed, we define
\ba\label{defcl2}
{\cal K}[C,R] &:=& \frac{d_R}{|N_C|} \sum_{D} \chi^R(D) \, {\cal K}[C,D]\,,
\ea
where $R$ denotes a unitary irreducible representation of the stabilizer group $N_C$, $\chi^R$ the corresponding character, and $d_R$ its dimension. We name these newly defined operators ${\cal K}[C,R]$  `charge ribbon operators'. They are projectors:
\be
{\cal K}[C , R]\,  {\cal K}[C' , R' ]  = \delta_{C , C'} \delta_{R, R'} {\cal K}[C , R] \; .
\ee

In summary, we have obtained closed ribbon operators ${\cal K}[C,R]$ which measure the excitation content of the region enclosed by the ribbon. This excitation content is characterized by two parameters: a conjugacy class $C$ of ${\cal G}$ and an irreducible representation $R$ of the stabilizer group $N_C$. The conjugacy class $C$ describes the curvature excitations, whereas $R$ measures the torsion. 
In more concrete physical terms in the gravitational context, we can understand the punctures as point particles coupled to $(2+1)$ gravity. In this case, $C$ encodes the mass of the particles, while $R$ the component of its spin (projected along the internal direction defined by the curvature). For Yang--Mills theories, on the other hand, $C$ is a measure of magnetic flux, while $R$ can be seen to measure the integrated flow of electric field into the region enclosed by the ribbon. This is because the translation operators are exponentiated versions of what would be an electric flux operator if $\mathcal G$ were a Lie group.

\subsubsection{ Fusion basis \label{subsub_fusion}}

A set of closed ribbon operators  $\{{\cal K}_\beta[C , R]\}$ is mutually commuting as long as the ribbons do not cross each other.  The fusion basis diagonalizes exactly a certain choice of such mutually commuting closed ribbon operators. This leads to a hierarchical set of ribbons. Indeed, first we consider the set of ribbons around the basic faces (or punctures), excluding the root face.  These define the basic excitations. One then fuses two excitations by considering ribbons around fused faces or punctures. In each step one fuses two excitations, which can be either basic ones or excitations resulting themselves from a fusion. One proceeds until there is only the outer face or root puncture left. Since we consider a sphere the ribbon around the root puncture agrees with the ribbon around the remaining punctures modulo orientation. Notice also that the faces do not have to be neighbouring with respect to a particular choice of underlying graph (this is why it is more powerful to get rid of the graph as we advocated above), but it is important that in the final set of closed ribbons, no two ribbons should cross each other. The choice of fusion scheme can be encoded in a fusion tree, where the end vertices of the tree correspond to  the end nodes of the graph, that is to its faces or punctures, and the root of the tree correspnds  the root of the graph, that is to its outer face. The trivalent vertices of the tree encode the fusion of two excitations into a new one. The edges of the fusion tree are labeled by pairs $(C_\beta,R_\beta)$  where $\beta\in\{1,\cdots, 2p-3\}$. These labels determine a set of fusion basis states, namely those fusion basis states the closed ribbon operators ${\cal K}_\beta[ C_\beta,R_\beta]$ project onto. 

Let us now construct explicitly such fusion basis states. We start with a holonomy basis state $\psi_{G,H}$ whose support is a two-punctured sphere $\S_2$ or equivalently a graph with two faces. Here the basis state $\psi_{G,H}$ is labeled by the pair $(G,H) \in \mathcal{G}\times \mathcal{G}$.
We explained in section \ref{sec_splitting}, that the gluing procedure for wave functions on $\S_2$ reproduces  the multiplication map of a well-known algebraic structure, namely the Drinfel'd double $\mathcal{D}(\mathcal{G})$ of the group $\mathcal{G}$.  In particular the Drinfel'd double admits a basis $\{[G,H]\}$ labeled by couples $(G,H) \in  \mathcal{G}^2$ ({\it cf.} appendix \ref{app_drinfeld} for a brief review). 
One can then define elementary excitations by demanding that the corresponding wave functions are stable under such a gluing procedure \cite{Lan2013}. This leads to the identification of elementary excitations with the irreducible representations of the Drinfel'd double $\mD(\mG)$, whose construction we now briefly briefly summarize  \cite{Koornwinder1999,Dijkgraaf1989,Dijkgraaf1991}. Following an induced-representation type of construction, one can show that the irreducible representations $\rho$ of ${\cal D}({\cal G})$ are labeled by couples $\rho=(C,R)$, where---as above---$C$ is a conjugacy class of ${\cal G}$ and $R$ an irreducible representation of the stabilizer group $N_C$.  The vector space $V_\rho$ on which the representation $\rho=(C,R)$ acts admits a basis labeled by $I=(i,M)$, with $i\in\{1,\ldots, |Q_C|\}$, where $Q_C={\cal G}/N_C$, and $M\in\{1,\ldots d_R\}$ the label of a basis of the representation space $V_R$. Within the representation $\rho$, we denote the matrix elements of a generic element $a=\left(\sum_{G,H}d_{G,H}[G,H]\right)\in {\cal D}({\cal G})$ by $D^\rho_{I'I}(a)$. The explicit expression of these matrix elements for the basis element $[G,H]$ is provided in appendix \ref{app_drinfeld}.

Finally, we define the fusion basis states $\psi_{\mathfrak{f}}$ on the two--punctured sphere $\S_2$ via the following generalized Fourier transform
\be
	\psi_{\mathfrak f} [\rho,I'I]\,=\,\frac{1}{|\mG|} \sum_{G,H} \sqrt{d_\rho}\;D^\rho_{I'I}([G,H]) \; \psi_{G,H}  \; .
\ee
As $\S_2$ is topologically equivalent to a cylinder we introduce the following graphical notation
\be
	\psi_{\mathfrak{f}}[\rho,I'I] = \begin{array}{c}\includegraphics[scale =1]{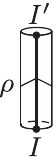}\end{array} \q \q \text{and} \q \q 
	\delta_{I'I} = \begin{array}{c}\includegraphics[scale =1]{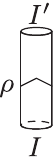}\end{array} \,,
\ee
where the second diagram represents the identity in the representation space $\rho$ of the Drinfel'd Double algebra. 

It is not hard to show at this point \cite{DDR1} that the closed charge ribbon operators ${\cal K}[C,R]$ project exactly onto these basis states $\psi_{\mathfrak{f}}[\rho,I'I]$ :
\be
\mathcal{K}[\rho_o] \psi_{\frak f}[\rho,I' I] = \delta_{\rho_o,\rho} \psi_{\frak f}[\rho,I' I].
\ee
We see, however, that these operators $\mathcal{K}[C,R]$ do not suffice to fully characterize the fusion basis, because of the presence of further basis label associated to the punctures ({\it i.e.} $I$ and $I'$) . One can introduce projection operators ${\mathbb P}_\alpha[\rho,I']$, whose purpose it is to project onto a fusion basis state carrying the label $(\rho,I')$ at the puncture $\alpha$. Therefore, the closed ribbon operators, together with these projection operators, give a complete set of commuting operators, characterizing the fusion basis.

With the irreducible representations at hand, we can make the notion of fusion of the basisc excitations more precise. As we have seen, the elementary excitations are described by irreducible representations $\rho$ of the Drinfel'd double, so that the fusion of two excitations is described by its recoupling theory:
\ba
\rho_1 \otimes \rho_2 = \bigoplus_{\rho_3}N^{\rho_3}_{\rho_1 \rho_2}\, \, \rho_3 .
\ea
For notational convenience, we will assume that $N^{\rho_3}_{\rho_1 \rho_2}$ either vanishes or is equal to 1 ({\it i.e.} that the irreducible representations of ${\cal D}({\cal G})$ form a multiplicity free fusion category), however the following derivations still hold without this assumption.  With a choice of basis (and phases) for the representation spaces, the decomposition is described by Clebsch--Gordan coefficients satisfying
\be
|\rho_3,I_3\rangle \,=\,  \sum_{I_1,I_2}  {\cal C}^{\rho_1 \rho_2 \rho_3}_{I_1I_2I_3}\,\,  	| \rho_1, I_1 \rangle \, \otimes \, | \rho_2, I_2 \rangle ,
\ee
and which can be graphically represented as 
\be
	\mathcal{C}^{\rho_1 \rho_2 \rho_3}_{I_1I_2I_3} = 
	\begin{array}{c}\includegraphics[scale =1]{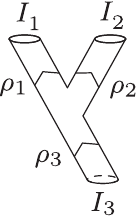}\end{array}  .
\ee
Relaxing the multiplicity-free assumption would lead to an additional multiplicity index for the Clebsch-Gordan coefficients.

We have now all the ingredients to define the fusion basis for the general case. Let $\psi_{\{G_{\alpha},H_{\alpha}\}_{\alpha}}$ be a holonomy basis state whose support is a $p$-punctured sphere. Since the number of punctures is $p$, the basis state is labeled by $(p-1)$ pairs $(G_{\alpha},H_{\alpha}) \in \mG^2$. Each one of these pairs corresponds to a basis excitation and can be thought as labelling a cylinder state $\psi_{G_{\alpha},H_{\alpha}}$. To define the fusion basis state, we first need to perform the transformation to the $[\rho',I'I]$-picture on each of these $(p-1)$ pairs respectively associated to $(p-1)$ punctures:
\ba\label{fusionb17}
\psi_{ \{ \rho_\alpha, I'_\alpha I_\alpha \}} \,:= \, \frac{1}{|\mG|^{p-1}} \sum_{\{G_\alpha,H_\alpha\}}  \, \psi_{\{G_\alpha,H_\alpha\}} \prod_{\alpha} \sqrt{d_{\rho_\alpha}} D^{\rho_\alpha}_{I'_\alpha I_\alpha}([G_\alpha,H_\alpha])  \; .
\ea
Here the index $I'_\alpha$ is associated to the corresponding (non--root)  puncture $\alpha$, whereas the index set $\{I_\alpha\}_\alpha$ are  all associated to the root puncture. Thus the root puncture carries the tensor product over all   representations $\rho_\alpha$. To make the fusion explicit, we decompose this tensor product by using a recoupling (or fusion) scheme encoded in a choice of  fusion tree.  That is, for every three--valent vertex  of the fusion tree we apply a Clebsch--Gordan coefficient to the states (\ref{fusionb17}). The contraction of all Clebsch--Gordan coefficients according to the fusion tree leads to a tensor ${\cal C}^{ \{\rho_\beta\} }_{ \{I_\alpha\}, I_{\rm r}}$.  This defines the fusion basis:
\ba
\psi^{\S_p}_{\mathfrak{f}}{ [\{\rho_{\beta}\}_{\beta=1}^{2p-3},\{I_{\alpha}'\}_{\alpha =1}^{p-1},I_{\rm r}]}
\,:=\,  \frac{1}{|\mG|^{p-1}} \sum_{I_{\alpha}}\sum_{\{G_\alpha,H_\alpha\}}  \, \psi_{\{G_\alpha,H_\alpha\}} \left(\prod_{\alpha} \sqrt{d_{\rho_\alpha}} D^{\rho_\alpha}_{I'_\alpha I_\alpha}([G_\alpha,H_\alpha])\right) \,    {\cal C}^{ \{\rho_\beta\} }_{ \{I_\alpha\}, I_{\rm r}} \; .
\ea
Equivalently, we can identify  (\ref{fusionb17})  with a product of states on the cylinder and write the fusion basis as
\begin{align}
	\psi_{\mathfrak{f}}^{\S_p}[\{\rho_{\beta}\}_{\beta=1}^{2p-3},\{I_{\alpha}'\}_{\alpha =1}^{p-1},I_{\rm r}]
	=  \sum_{I_{\alpha}}\,\, \prod_{r=1}^{p-1}\psi_{\mathfrak{f}}[\rho_{\alpha},I_{\alpha}'I_{\alpha}]  \,\, {\cal C}^{ \{\rho_\beta\} }_{ \{I_\alpha\}, I_{\rm r}} \; .
	\label{eq_fusbasis_cyls}
\end{align}
This leads to the following encoding of the fusion basis in a  graphical representation, which also includes the fusion tree:
\be
	\psi_{\mathfrak{f}}^{\S_p}[\{\rho_{\beta}\}_{\beta=1}^{2p-3},\{I_{\alpha}'\}_{\alpha =1}^{p-1},I_r]= 
	\begin{array}{c}\includegraphics[scale =1]{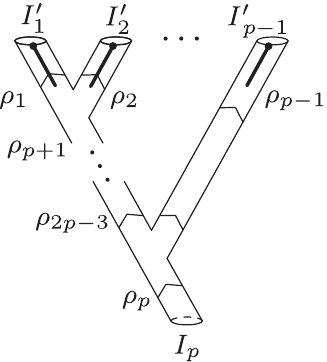}\end{array}  \q .
\ee
Here the $(p-1)$ upper cylinders are associated to the $(p-1)$ punctures and are connected  to each other via Clebsch-Gordan coefficients.
This basis can be shown to be orthonormal and complete \cite{DDR1}, and to diagonalizes the closed ribbon operators supported on the loops associated with the relevant fusion tree:
\be
\mathcal{K}_\beta[\rho]\psi_{\mathfrak{f}}^{\S_p}
= \delta_{\rho,\rho_\beta}\psi_{\mathfrak{f}}^{\S_p}
.
\ee

Moreover, with the above  graphical notation, it is clear that the usual lattice-based representation has been abandoned in favour of a representation relying exclusively on the excitations and the way they fuse together. Therefore it is natural at this point to define a region not so much in terms of the underlying lattice but only in terms of the excitations it contains.

 Note that a different choice of fusion tree would lead to different fusion basis. The fusion basis given above is based on a particular such choice of tree. In the following, see also appendix \ref{app_alt}, other choices will turn out to be more relevant. 

\subsubsection{Fully gauge-invariant wave functions}

Fully gauge-invariant wave functions can be also easily described in terms of the fusion basis.  The gauge invariant projection at the puncture $\alpha$, implies that it carries a trivial representation labels $R_\alpha=0$. This trivial representation label $R_\alpha=0$, however, still allows the index $I_\alpha =(i, M\equiv0)$ to range among the values of $i$ labeling the elements of the quotient $Q_{C_\alpha}={\cal G}/N_{C_\alpha}$. The gauge-invariant projection induces, however, an averaging over the elements of $Q_{C_\alpha}$, labeled by $i$. Therefore, the gauge-invariant projection of a fusion-basis states is effectively labeled only by the conjugacy classes $C_\alpha$. Note, however, that in the non-Abelian case fully gauge-invariant basis states can actually be labeled by $\rho_\beta=(C_\beta,R_\beta)$ which include non--trivial representations $R_\beta$ for $\beta$'s associated to inner edges of the fusion tree.

\subsubsection{Open ribbon operators} 

\begin{figure}[h]
	\includegraphics[scale =1]{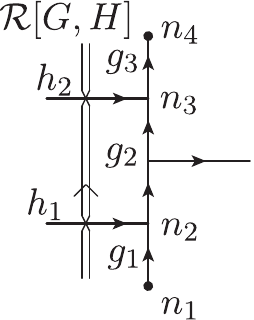} \q 
	\caption{Action of an open ribbon operator from nodes $n_1$ to $n_4$. The ribbon acts as a Wilson path operator on the holonomy parallel to the ribbon {\it {\it i.e.}} $g_3g_2g_1$ and as a translation operator on the holonomies crossed by the ribbon {\it {\it i.e.}} $h_2$ and $h_1$. }
	\label{fig_open}
\end{figure}

We have already introduced the closed ribbon operators, which measure the excitations without changing the excitation content.  We now introduce open ribbons which create pairs of excitations (or change the pre-existing excitations) at their ends. To keep the excitations confined at the punctures, open ribbons operators have to start and end there. 
 To fix the definition of open ribbon operators, we fix the conventions that they act on the graph links to their right. Eventually, one can show that---as for the closed ribbon operators---the action does not dependent on the choice of underlying graph but only on the relative position of the ribbon with respect to the punctures (see {\it e.g.} \cite{DDR1}).
To keep matter simple, we define the open ribbon operators only via the example of figure \ref{fig_open}, and refer for the general cases to \cite{DDR1}.

The ribbon operator ${\cal R}[G,H]$ is parametrized by $(G,H) \in {\cal G}^2$ and combines (as before) a holonomy operator part, acting on the holonomy parallel to the ribbons, and a translation operator part, acting on the holonomies crossed by the ribbon:
\ba\label{oribbon1}
\left({\cal R}[G,H] \psi\right) (g'_1,g'_2,g'_3,h_1,h_2, \ldots)
\,=\, \delta( G, g'_3 g'_2 g'_1) \,\, \psi( g'_1,g'_2,g'_3,   (g'_3 g'_2)^{-1} H^{-1} g'_3 g'_2 h_1,  (g'_3)^{-1} H^{-1} g'_3 h_2 , \ldots)  \;.
\ea
The parallel transport for the translation part  insures that  gauge invariance is preserved at the internal nodes.  Furthermore, the action is defined such that the curvature is changed only for the faces that include the source and target end-nodes  of the ribbon. 

Let us define a vacuum state as being a state without excitations, that is, the curvature is vanishing for all faces (or punctures). Vanishing torsion, on the other hand, means that the states are completely spread over the holonomies going from one puncture to another. Thus the (BF) vacuum state is given in the holonomy basis and up to normalizations by
\ba
	\psi_0 \sim \sum_{G_\alpha} \psi_{ \{G_\alpha, H_\alpha=e\}} \; ,
\ea
where $e$ denotes the unit element of ${\cal G}$. 
Applying a ribbon operator to the vacuum state, we create curvature and torsion excitations at its ends. Because we can create such excitations only in pairs, they are of a quasi--local nature. We can also ask for charge ribbon operators that would create basic excitations, labeled by the Drinfeld double algebra representations $\rho=(C,R)$. In fact, applying the generalized Fourier transform, we obtain the operators
\ba
	\mR[\rho,I' I] = \frac{1}{|\mathcal{G}|}\sum_{G,H}\sqrt{d_{\rho}}\mR[G,H] D^{\rho}_{I' I}([G,H]),
	\label{chargeribbon}
\ea
which indeed generate the fusion basis for the two--punctured sphere from the vacuum state
\ba
	\label{ribbonstate}
	\psi_{\mathfrak{f}} [\rho,I'I] = |\mG|^{3/2} (\mathcal{R}[\rho,I',I]\psi_0)(g_1,\cdots,g_4)  \equiv
	\begin{array}{c}\includegraphics[scale =1]{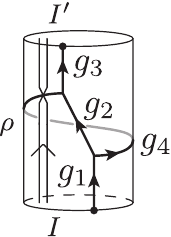}\end{array}
	\; .
\ea
 From the relation between the fusion basis and the cylinder states given in equation \eqref{eq_fusbasis_cyls} it should be evident that by appropriately acting with these charged ribbon operators on the  vacuum, one can indeed generate the whole fusion basis.
An important related fact is that it is possible to define a ``lengthwise'' product of open ribbon operators---appropriately sharing a start and end-puncture---which leads to a new open ribbon operators supported on the combined path. This multiplication reflects the gluing operation and satisfies the same Drinfel'd double algebra (see \cite{DDR1} for details).


\section{Entanglement entropy in lattice gauge theories \label{sec_ent_lat}}

Having introduced the fusion basis, we now have all the ingredients to define a new notion of  entanglement entropy through the procedure of section \ref{sec_ent}. There, we assumed to have a basis labeled by representation indices $\{\rho\}$, and that the system under scrutiny was divided in two regions generically associated to a set of representation labels $\{\rho_\partial\}$.  Now, in the case of the fusion basis, and assuming that there is only one connected boundary, this set of labels can always be reduced so that it includes {\it only one} representation $\rho_\partial$.

This representation label describes the outcomes of the closed ribbon operator going along the boundary between regions.  The extension procedure described in section \ref{sec_splitting}, introduces an extended Hilbert space ${\cal H}_{\rm ext}={\cal H}_A \otimes {\cal H}_B$, which factorizes into two Hilbert spaces. These two Hilbert spaces correspond to the two systems one obtains after splitting the surface $\Sigma$ along the boundary, see the discussion in section \ref{sec_splitting}. 

\begin{figure}[h]
	\includegraphics[scale =1]{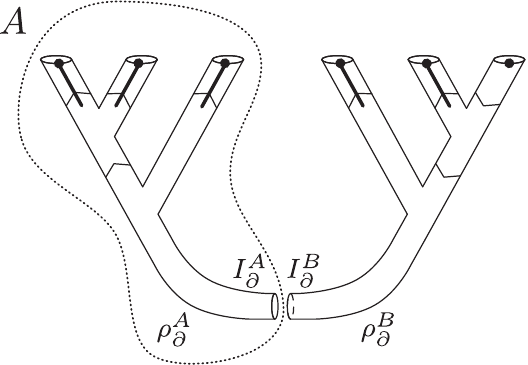} 
	\caption{For a given fusion tree we identify the region $A$ as a set of punctures and $B$ its complement. The splitting is performed by cutting a cylinder of the fusion tree. This cut requires the introduction of additional punctures. Note that the fusion tree employed here is the same as the one appearing in the alternative fusion states defined in Appendix \ref{app_alt}.}\label{FigCut1}
\end{figure}

Figure \ref{FigCut1} represents the cut into two regions in the picture using punctures and the fusion tree. In the usual lattice picture the cut proceeds along the boundary of the plaquettes. More precisely we can imagine to double the Wilson loop around {\it e.g.} the region $A$ into two loops which go closely parallel to each other. The two Wilson loops are connected with one `small' link, whereas the area between the two loops carries flat connection.  (Demanding flat connection for this area and gauge invariance at the additional nodes, we can uniquely map the state on the original graph to the graph with the doubled Wilson loop.) The cut proceeds then in-between the two Wilson loop and cuts the link connecting the two loops. 
\begin{figure}[h]
	\includegraphics[scale =1.2]{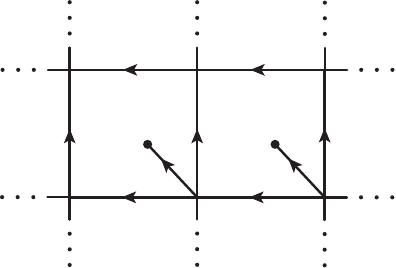} \q  \q 
	\includegraphics[scale =1.2]{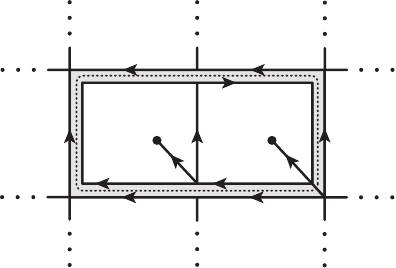} 
	\caption{ Example of splitting on a square lattice between two plaquettes and the complementary graph. First the plaquettes are isolated by doubling the boundary links. An extra link is added to connect the isolated plaquettes to the rest of the lattice. The group element on this extra link is determined by the flatness constraint imposed on the face represented in gray. The splitting is finally performed along the dashed line. }
\label{fig_latticesplit}
\end{figure}

Let us mention some key differences between this extension procedure and the one which uses spin network states as in \cite{Donnelly2008,Donnelly2011,Donnelly2014}.  The use of the fusion basis emphasizes the role of the excitations with respect to the BF vacuum, which in turn prescribes gauge-invariant flat connection. In $(2+1)$ dimensions this means that only the position of the punctures matters, not the graph itself.  This feature is particularly important for the application to $(2+1)$ dimensional gravity, where the BF vacuum, with defect excitations describing particles, give the physical states \cite{noui2004,noui2006,Meusburger2008}. 

The two procedures can be also compared in how `big' the extension of the Hilbert spaces is. This can be quantified in terms of how many constraints are violated in the extended Hilbert space. In the case of the spin network basis this includes all the Gau\ss~constraints at the two--valent nodes that result from the boundary cutting links. In the case of the fusion basis this includes only one Gau\ss~constraint and one flatness constraint. In the picture described above the Gau\ss~constraint needs not to hold anymore for the link which is cut into two. And the flatness constraint that is violated in the extended Hilbert space is the one between the two Wilson loops arising from doubling the Wilson loop along the boundary.

This is, nonetheless, a sort of minimal choice---one that can, importantly, always be made. A more general splitting can be introduced also for the fusion basis. This corresponds to choosing somewhat arbitrary more marked points along the boundary of the two regions. In this way, more Gau\ss~constraint violations will be added (but no extra curvature violation, see \cite{DDR1}). From the observable algebra perspective, this corresponds to declaring observable a series of gauge-variant holonomies along a set of paths which partition the boundary. This extension will only introduce additional vector-space indices ({\it i.e.} new indices `next to' the $(I,I')$ in $V_{\rho_\partial}$). Later, we will briefly discuss what kind of consequence this has for the entanglement entropy.

We are now going to explicitly calculate the entanglement entropy for some simple choices of states and regions.
To remind the reader, the entanglement entropy has three contributions 
\ba\label{ent2ex}
S_A\,=\, H( P(\rho_\partial)) \,+\,  \langle \ln \dim \rho_\partial \rangle \,+\, \langle S_A( \D^A(\rho_\partial)) \rangle ,
\ea
where $P(\rho_\partial)$ is the classical probability distribution 
\ba
P(  \rho_{\partial}) = \sum_{\rho_A, I_A; \rho_B, I_B} \psi( \{\rho;I\}) \overline{  \psi( \{\rho;I\})   } , 
\ea
while $\langle {\scriptsize \bullet} \rangle$  and $H(P(\rho_\partial))$ denotes the average with respect to $P(\rho_\partial)$ and its Shannon entropy respectively.

\subsection{Fusion basis states (and BF vacuum)}

We start with  a fusion basis state  on the $p$--punctured sphere ${\S}_p$. Its expansion in the fusion basis of course gives
\ba
\Big( \psi_{\frak f}{[ \{\rho^0;I^0\}]}\Big)( \{\rho;I\})\,=\, \prod_\beta \delta_{\rho_\beta,\rho^0_\beta} \prod_\alpha \delta_{I_\alpha,I^0_\alpha}\, .
\ea
 Here $\beta$ labels the edges of the fusion tree and $\alpha$ its endpoints.  We partition the punctures into two sets $A$ and $B$.  We choose a fusion tree such that the $A$ and $B$ sets are only connected by one fusion tree edge labeled by $\rho_\partial=\rho_\partial^0$.  That is we are only considering fusion basis states from a basis characterized by a tree which is `compatible' with the prescribed splitting. Such a basis always exists (and in fact, there are many).

In this case the classical probability distribution $P(\rho_\partial)$ is peaked on one particular value $P(\rho_\partial)=\delta( \rho_\partial, \rho_\partial^0)$, thus the associated Shannon entropy vanishes. The density matrices  $\D^A( \rho_\partial)$ are defined to vanish for $\rho_\partial \neq \rho_\partial^0$, and the density matrix $\D^A( \rho_\partial^0)$ has only one non--vanishing entry equal to $1$ on the diagonal, and therefore give no contribution to the entanglement entropy (\ref{ent2ex}).  Therefore, we are left with the middle term in (\ref{ent2ex})
\ba\label{family1}
S_A(
 \psi_{\frak f}{[ \{\rho^0;I^0\}]}
) \,=\ \ln \dim \rho_\partial^0 .
\ea
Notice that we find a vanishing entropy for the BF vacuum state, as in this case $\dim \rho_\partial^0 =1$. This agrees with the result found  in \cite{Casini2013} for Abelian gauge theories with the magnetic centre choice on the BF vacuum state.  For Abelian structure groups, $\rho=(C,R)$ is labeled by a group element (as $C = \{g\}$) and an irreducible representation of its stabilizer, that is of the whole group. This irreducible representation is---the group being Abelian---one-dimensional. Hence,  in this case $\dim \rho = (\dim C )(\dim V_R )= 1$, and the entanglement entropy vanishes for  (compatible) fusion basis states.

We can also consider gauge invariant projections of fusion basis states, as defined in section \ref{sec_fusion}.  We can consider these states both in the Hilbert space ${\cal H}_p$ which allows for torsion excitations at the punctures, and its gauge invariant projection, where torsion excitations do not appear for the punctures (but can---in the non-Abelian case---appear for internal edges of the fusion tree). In both cases the result is the same as in (\ref{family1}). The only difference could have been in the contributions from the density matrices $\D^A( \rho_\partial)$, but these do describe pure states also after the action of gauge-averaging projectors. This is because such projectors act locally within one single region. 

Let us shortly come to the extension mentioned at the end of this section's introduction. This extension consists in a generalization of the Hilbert space ${\cal H}_p$ and of the corresponding fusion basis to the case where more than one link is allowed to end on a given puncture (in this case the puncture that is identified with the boundary between the two regions, see \cite{DDR1}). It turns out that the corresponding fusion basis is still labeled with the same representations, the only difference is in  the vector space indices $I$ at the extended puncture: for each additional marked point accompanied with a link ending at the puncture the index range is multiplied by $|{\cal G}|$, the order of the group. 
Associated to this generalization of the fusion basis we can also consider a generalization of the gluing and extension procedure. Basically we can decide by how many graph links we wish to connect region $A$ and region $B$.
 Note that this reintroduces a graph dependence\footnote{More precisely one decides on the number of marked points along the boundary through which the crossing links have to pass.} that previously we fixed by using an equivalence relation between states, that allowed us to change the underlying graphs. This enabled us to always reduce to the case that region $A$ and region $B$ are connected by only one link. 

Using this generalized procedure the adjustment of the entropy formula is very simple. The result is that for each additional marked point the entanglement entropy increases by $\ln|\mathcal G|$. Interestingly, it turns out that the following relations between dimensions hold (see {\it e.g.} \cite{DDR1})
\be
\dim {\cal D}(\mG) = |\mG|^2 = \sum_\rho (\dim \rho_{{\cal D}(\mG)})^2,
\ee
where the first is a dimension of the Drinfel'd double seen as a vector space spanned by the basis $\{[G,H]\}$, the second is the cardinality (order) of the group $\mG$, and the last one is again a dimension of a vector space,  {\it i.e.} of $V_\rho$. The last term in the above equality is also known as the (square of the) `total quantum dimension' of the fusion category given by the irreducible representations of ${\cal D}(\mG)$. Due to the lack of fonts for a capital `D', we will call this quantity, {\it i.e.} the total quantum dimension of ${\cal D}(\mG)$, simply $\Omega$:
\be
\Omega_{{\cal D}(\mG)}  := \sqrt{\sum_\rho (\dim \rho_{{\cal D}(\mG)})^2~}.
\ee
Hence, we find that for a (compatible) extended fusion basis state, with $m$ marked points\footnote{This number, $m$, includes also the marked point which is always supposed to be there. Hence $m\geq1$ in this definition, and $m=1$ for `minimal' fusion basis states.} at the boundary puncture, itself labeled by $\rho_\partial$, the entanglement entropy amounts to
\be\label{more1}
S_A(
 \psi_{\frak f}^m{[ \{\rho^0;I^0\}]}
) \,=\ \ln \dim \rho_\partial^0 + (m-1)\ln \Omega_{{\cal D}(\mG)}
\ee
Notice, that by using such a graph-dependent formula, one obtains also a non--vanishing contribution for the BF vacuum state. 
This can be understood by realizing that the extended Hilbert space allows now a refined information on the gauge connection along the boundary. For example, with two links crossing the boundary we can specify the holonomy between the corresponding  two marked points on the boundary. This holonomy can be non--trivial even if the holonomy along the complete boundary is.

Remarkably, the result (\ref{more1}), applied to the BF vacuum with $\mathbb Z_2$ gauge group, does agree with the entanglement entropy defined via  the Hilbert space extension based on the spin network basis (or with the electric centre choice) found in \cite{Donnelly2011}.  To this end one has to choose $m$,  the number of links connecting regions $A$ and $B$ to agree in both procedures. As we pointed out however, with our procedure, based on the fusion basis we are free to perform the (BF representation based) continuum limit keeping $m$ fixed, ensuring a finite (or vanishing if $m=1$) entanglement entropy for the BF vacuum. In contrast, using the Ashtekar--Lewandowski representation, the continuum limit of the BF vacuum state requires to take $m$ to infinity, thus leading to an infinite entropy in this limit. Nevertheless this agreement in results is striking and it would be interesting to see if this holds for more generic states.

\subsection{States generated by the action of open charge ribbon operators \label{subsec_calc}}

Let us go back to the usual Hilbert space ${\cal H}_p$ and consider 
another class of examples, namely states that are generated from the BF vacuum by applying a number of open ribbon operators, that are going from region $B$ to region $A$.  We start with the case of two charge ribbon operators ${\cal R}_1[\rho_1]$  and ${\cal R}_2[\rho_2]$, associated to two different paths. Thus we have to consider  states on the 4--punctured sphere $\S_4$.

\begin{figure}[h]
	\includegraphics[scale =1]{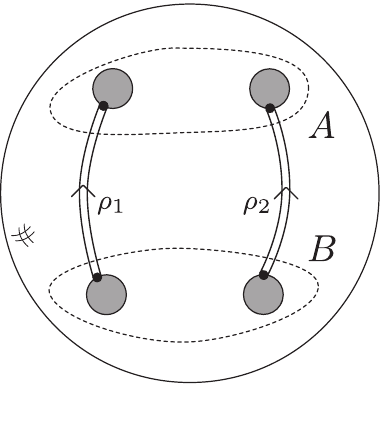} 
	\caption{A class of states is generated from the BF vacuum by acting with several open ribbon operators going from a region $B$ to a region $A$. The entanglement entropy for such states can be computed using the fusion basis construction.}
	\label{fig:4punctribbons}
\end{figure}

In this case too, the contribution  $S_A( \D^A ( \rho_\partial))$ vanishes as $\D^A( \rho_\partial)$  describes again  a pure state. 
As shown in more detail in appendix \ref{app_thecalcul} the other two contributions to the entanglement entropy are determined by the probability distribution 
\ba
P(\rho_\partial)\,=\,   N^{\rho_\partial}_{\rho_1 \rho_2}  \, \frac{ \dim \rho_\partial} {\dim \rho_1 \, \dim \rho_2}  .
\ea
Note that this agrees with the expectation value of the closed ribbon operator ${\cal K}[\rho_\partial]$ along the boundary of the two regions, {\it i.e.}  $P(\rho_\partial) = \langle \mathcal{K}[\rho_\partial] \rangle$.
One has indeed $\sum_{\rho_\partial} P(\rho_\partial)=1$. With this probability distribution one can compute the entanglement entropy for the state under consideration to be
\ba
S_A\,=\, \ln \dim \rho_1 + \ln \dim \rho_2  .
\ea
More generally we can consider $n$ non--intersecting ribbons ${\cal R}_a[\rho_a]$, $a=1, \ldots n$, all going from region $B$ to region $A$.  These will generate a state proportional to the normalized state (see appendix \ref{app_thecalcul}) 
\ba
&&\sum_{\rho_{n+1}, \ldots, \rho_\partial}  N_{\rho_1\rho_2}^{\rho_{(n+1)}} N_{\rho_{(n+1)}\rho_3}^{\rho_{(n+2)}} \cdots  N_{\rho_{(2n-2)}\rho_{n}}^{\rho_{\partial}} \sqrt{ \frac{d_{\rho_\partial}} { d_{\rho_1} \cdots d_{\rho_n}}}
\nn\\
&&\q\q  \widehat{\psi}_{\mathfrak f}^{\mathbb{S}_{2n}} [ \{\rho_a\}, \{ \rho, \ldots, \rho\}_{(n+1)}^{2(n-1)}, \rho_\partial,  \{ \rho, \ldots, \rho \}_{(n+1)}^{2(n-1)}, \{\rho_a\}; \{ I_a\}, \{I'_a\} ]  ,\q\q
\ea
 where the states $\widehat{\psi}_{\mathfrak f}^{\mathbb{S}_{2n}}$ are the orthonormal fusion basis states defined in Appendix \ref{app_alt}.
Again the entanglement entropy is determined by the first two terms in (\ref{ent2ex}) to be  simply
\ba
\label{RibbEnt}
S_A\,=\, \sum_a  \ln \dim \rho_a  .
\ea
We thus get the entropy to scale with the number of ribbon operators crossing the boundary.    Note that we only get a non--vanishing entropy for these states due to the first two ``classical" contributions to the entanglement entropy (\ref{ent2ex}).  The distillable entanglement entropy would therefore be vanishing.  This is probably due to the fact that the ribbon operators generating these states are not part of the reduced observable algebra ${\cal O}_{\text{red}}$, that underlies the definition of distillable entanglement entropy. 

\subsection{Comparison with the literature}

A similar result to (\ref{RibbEnt}) was also obtained in \cite{Dong2008,Wen2016}, which considered the entanglement entropy for Chern--Simons theory.  Whereas \cite{Dong2008} uses the replica trick  in a covariant path integral approach combined with surgery techniques, \cite{Wen2016} employs  again the replica trick but within the conformal field theory induced on the boundary.  

In these two references, the entanglement entropy is also computed for states on the sphere as generated by the insertion of Wilson line operators. Notice, that the Chern--Simons Wilson lines involve holonomies of a Poisson non-commutative connection, which is therefore not the connection involved in our underlying states. At the same time, our states are generated by the action of ribbon operators, which are exactly Wilson lines for the double of the group. The analysis of \cite{MeusburgerWise,Meusburger2016} actually shows explicitly that the quantization of BF theory with gauge group $\mG$---like the one we are considering in this paper---is equivalent to the combinatorial quantization of Chern--Simons theory for the double ${\cal D}(\mG)$. This result\footnote{Their result holds `on-shell' as a quantization of the moduli space of flat connections. However their construction involves `off-shell' quantities as well, and the claim we are going to make can be in principle checked rigorously. We postpone the study of this interesting question to future work.} suggests that by appropriately identifying the structures in our computation and in theirs, they should match exactly, in the sense that they aim to compute the same physical quantities in two equivalent quantization schemes. After this premise, we can now compare the our results.

First of all, while \cite{Dong2008} finds finite answers, those of \cite{Wen2016}---found through computations in the edge field theory---are divergent. This divergence is due to an offset proportional to the central charge of the dual conformal field theory, and can therefore be thought as being associated to its zero-point energy. We will come back to this term later, for the moment let us focus on the other terms on which \cite{Dong2008} and \cite{Wen2016} agree.
These terms contain two contributions. The first contribution $S_1$ coincides with our result \eqref{RibbEnt}, while the second $S_2$ is another universal offset determined by the total quantum dimension $\Omega$
\be
S_2=-\ln \Omega.
\ee
This can be interpreted as a `vacuum contribution' to the entanglement entropy. It is a negative contribution.
As such, it cannot result from a finite Hilbert space computation, and in fact we do not find it.
(This term is also related to the so-called topological entropy \cite{Kitaev:2005dm,Levin2004}, whose study in the framework of \cite{Levin2004} needs the application of the fusion basis technique to disconnected regions. We postpone this study to future investigations.)

Finally, coming back to the (positive) divergent term found using the dual field theory, it is interesting to speculate about its possible origin in the discrete framework.
In Chern--Simons theory on a three-manifold with boundaries, the dual CFT living on the boundary is given by the WZNW model. If the three-manifold is a two-disk times an interval, a way to see the appearance of the WZNW model is by solving explicitly the flatness constraint in the bulk of the disk. In the standard notation $A_\mu(x) = g(x)^{-1}\partial_\mu g(x)$.  This leaves us with the field $g(x)$ on the boundary, since its bulk contributions to the action essentially cancel out (modulo topological terms). Therefore, it is the choice of local frame $g(x)$ in which the flat gauge connection is evaluated which encodes the boundary CFT field (see also \cite{Carlip:2005zn} for these derivations in relation to three-dimensional gravity). At this point, it is natural to draw a parallel between the frame $g(x)$ in the continuum theory and the local frames we introduce in the refined boundary-puncture picture. If this is done, it is clear that infinitely many refining points on the boundary-puncture are needed to fully capture the dual field theory. In this limit also our entropy diverges. But as our procedure and the regularization procedure used in \cite{Wen2016} are completely different, a more precise relation is difficult to obtain at this stage. In spite of this, we find this question extremely interesting.

\subsection{TQFT based continuum limits }

In section \ref{sec_fusion}, we defined the Hilbert spaces ${\cal H}_p$ that capture  $2p-2$ degrees  of freedom of a gauge theory on a fixed graph or lattice  embedded on the 2-sphere.  The set-up taken here allows to embed these Hilbert spaces ${\cal H}_p$ into a continuum Hilbert space ${\cal H}_{\text{cont}}$. 

This leads to the so-called BF representation \cite{DGflux,DGfluxC,DGfluxQ}, consisting of a Hilbert space which supports a representation of a continuum observable algebra, formed by the ribbon operators. The Hilbert space is based on the BF vacuum state, which is sharply peaked on flat connections, and is spanned by states that arise from the action of finitely many open ribbon operators on this vacuum state. The ribbon operators are then allowed to end at arbitrary points, which hence define the punctures. Such a Hilbert space can be constructed as an inductive limit from a family of Hilbert spaces based on fixed graphs (or more precisely equivalence classes of graphs, see \cite{DGflux,DGfluxC,DGfluxQ}).  In this latter viewpoint one puts all degrees of freedom, which are finer than the ones supported by the fixed graph, in the BF-vacuum state. 

This leads us to the following interpretation of the result (\ref{RibbEnt}). 
First of all, the vacuum state has a vanishing entanglement entropy. 
Then, each (charge ribbon) operator ${\cal R}[\rho]$ that connects  the two regions contributes to the entanglement entropy with $\ln \dim \rho$.
This is despite the fact that in our definition of entanglement entropy we make explicit use of a particular fusion basis, which involves only one ribbon crossing the boundary.

An analogous result holds for the `electric centre' choice, or its spin-network based extension \cite{Donnelly2014}.  Here we can also introduce a continuum Hilbert space, known as Ashtekar--Lewandowski representation \cite{Ashtekar1986,Ashtekar1991,Ashtekar1993}. This Hilbert space is based on a vacuum state peaked on vanishing electric fluxes. Wilson loops and lines do now act as creation operators. In fact a spin-network state results from the action of a network of Wilson lines connected via intertwiners. Using the spin-network based extension, one also finds that the entanglement entropy of a spin-network basis state is given by $\sum_a \ln \dim \rho_a$ where now $\rho_a$ denotes the irreducible representation of  ${\cal G}$ associated to the $a$-th spin-network link which crosses the boundary.  Again, one finds that the associated vacuum (the Ashtekar-Lewandowski vaucuum) has vanishing entanglement entropy. 

However, if we express the BF vacuum in the spin-network basis and use the procedure of \cite{Donnelly2014} to define the entanglement entropy, we notice that the result depends on the underlying graph. In particular, taking a refinement limit for the graph we would find a divergent result.  Such a refining limit is necessary in this case to fully ({\it i.e.} everywhere) describe the BF vacuum. In other words, the BF vacuum is an infinitely excited state with respect to the Ashtekar--Lewandowski one, and the entanglement entropy reflects this fact. On the other hand, using the fusion basis and the related Hilbert space extension we emphasize the excitations relative to the BF vacuum itself. This leads to a result which is graph independent, a fact that makes our method applicable to the case of $(2+1)$ dimensional gravity, which is described via a BF theory with defects describing point particles.  Of course, attempting a description of the Ashtekar--Lewandowski vacuum (that is the strong coupling limit of Yang Mills theory) in terms of the fusion basis would also lead to results which are graph dependent or divergent.

Thus, we see that different notions of entanglement entropy are also adjusted to different notions of representations, or phases, or regimes.  The BF representation corresponds to the (Yang Mills) weak coupling regime, and the Ashtekar--Lewandowski representation to the strong coupling regime. The excitations are in both cases (quasi--local) deviations from the weak coupling and strong coupling limit, respectively. In both cases, one deals with a topological theory with defect excitations.

We wish to emphasize that the vacua, both in the BF as well as in the Ashtekar--Lewandowski representation, describe theories without propagating degrees of freedom. That is, one can define Hamiltonians for which these vacua are the lowest energy states, which are moreover gapped. It is, thus, consistent to associate a vanishing entanglement entropy to these states. To describe the vacua of theories with propagating degrees of freedom in the continuum limit, we would need to introduce infinitely many excitations with respect to either of the vacua. This would lead to the usual divergent behaviour for the entanglement entropy in quantum field theories with propagating degrees of freedom.


\section{Entanglement entropy in gravity \label{sec_ent_gravity}}

The notion of entanglement entropy is usually (but not exclusively) associated to subsystems describing a region of space, which is specified by coordinates.  One would like to link such a  choice of  region to a  subset of observables, commuting with the remaining observables, but we have seen that this is already an ambiguous process for gauge systems.  These difficulties are much more enhanced in general relativity. 

In background independent theories, such as general relativity, regions specified by coordinates lack an a priori  operational meaning. Alternatively, one can define regions through matter or metric fields. This is very similar to  employing relational observables \cite{Rovelli2001, Giddings2005, Ditt04,Ditt05} as  gauge invariant observables in general relativity. Here the metric or matter fields are used as a reference system, in which other fields can be expressed in \cite{Ditt05,Kuchar1990,Kuchar1994}. Thus one can also attempt to specify `physical regions' by employing a `physical reference system'.   

Relational observables can be computed in an approximation scheme \cite{DittTambo1, DittTambo2}, which also allows an understanding  of how the standard observables of quantum field theory (on a fixed background) arise as approximations to fully gauge invariant observables. 
%
%
%
%

A crucial drawback of using matter or metric fields as reference system is that there will be phase space regions in which these fields are not suited as clocks and rods. In some systems smooth gauge invariant observables might not exist \cite{Chaos1, Chaos2}.  Thus one expects that notions of locality can be realized only for a certain class of states \cite{Giddings2015} and are furthermore only approximate \cite{Giddings2005,DonnellyGiddings2016}. 

Another key point is the question whether one can find a split of the observable algebra into mutually commuting sets describing (approximately local) subsystems \cite{Giddings2015,DonnellyGiddings2015,DonnellyGiddings2016}. For example, the approximation scheme developed in \cite{DittTambo1} regains the usual quantum field theoretical observables on a fixed background at lowest order, but  at higher orders it includes non-local terms. Giddings and Donnelly argue that observables creating {\it e.g.} matter fields, need to be gravitationally dressed, in order to capture the accompanying gravitational field. In contrast to Yang--Mills theories, this dressing cannot be screened and leads to an inherent  non-local structure \cite{DonnellyGiddings2016}. 

Using relational observables, one can deduce the commutator algebra by using Dirac brackets \cite{Ditt04}. Realistic (that is relativistic) matter fields allow, however, only for an approximate localization \cite{Giddings2005,DittTambo1}.  Physical coordinates built from geometry ({\it e.g.} by using geodesics) lead, at least so far, to non--local algebras \cite{DittTambo1,Bodendorfer2015}. 

The exploration of the diffeomorphism invariant observable algebra is very difficult, as it basically requires to understand and solve the dynamics of the system. This is, of course, a very challenging task for the four--dimensional theory.
On the other hand  three--dimensional general relativity is much simpler: it describes locally flat spacetimes (or homogeneously curved ones, in presence of a non-vanishing cosmological constant). This also means that one has no local degrees of freedom, but only global topological ones.  Introducing matter changes this situation, but it requires again a solution of the theory. To keep the system solvable, one can consider the coupling of point particles, which leads to a topological field theory with curvature and torsion defects, as discussed in this paper. 

In fact (Euclidean) 3D gravity without a cosmological constant can be described by a BF theory with $\SU(2)$ structure group. The coupling of point particles leads to curvature and torsion defects \cite{Hooft1993, deSousa1990,Freidel2004,noui2006}.  In short, the formalism needed to describe 3D gravity is very close to the formalism used here.  Moreover including a positive cosmological constant, one has to work with a  $\SU(2)_q$ structure group, with $q$ a root of unity. This leads to a finite dimensional Hilbert space, as in the finite group case we discussed. The fusion basis, ribbon operators, as well as gluing and cutting procedures are also available in the quantum group case \cite{DGTQFT}.

The fusion basis diagonalizes a maximally commuting subset of gauge invariant ({\it i.e.} Dirac) observables, given by closed ribbon operators.  A conjugated set of Dirac observables is provided by  open ribbon operators going from one particle to another.  Adopting the definition of entanglement entropy laid out here, a region is indeed specified by its matter content, that is by the particles contained in this region.  Note that we do not have to  specify the precise (geometric) position of the boundary, we only need to declare which particles belong to which regions. The geometric information is rather contained in the state under consideration.

We will refer to particles in `region' $A$ as $A$--particles and the remaining particles as $B$ particles. The associated $A$--observable algebra includes closed ribbons surrounding subsets of $A$--particles, with the exception of the closed ribbon surrounding all $A$--particles  (and therefore also surrounding all the $B$ particles as we consider spherical topology).  This closed ribbon forms the centre of the algebra in the language of \cite{Casini2013}. Additionally one can construct Dirac observables from open ribbon operators going from one $A$--particle to another $A$--particle. 

Ribbon operators crossing the boundary  cross also the closed ribbon along the boundary  and would therefore not commute with it. These operators cannot be associated to either the $A$ or $B$ region.  

Thus, although we can solve in this example the problem of how to define a region, we still have to modify the observable algebra, removing ribbons that cross from the $A$ to the $B$ regions from the operator algebra on which the entanglement entropy is being defined (if we follow the definition of \cite{Casini2013}).

The notion of subsystems for 3D gravity used here differs in key points from the proposal (so far on the classical level) of \cite{DonnFreid}. 
There one introduces additional fields, that allow to fix the boundary in terms of embedding or coordinate functions.  This has been motivated as a generalization of the extended Hilbert space construction (or rather its classical version). Here we point out that as there are different extension procedures in lattice gauge theories, this is also very likely to hold for gravity. The procedure laid out in this work can be applied to 3D gravity, and leads to much less extra structure compared to \cite{DonnFreid}. Furthermore, we can also state the definition of subsystems in terms of mutually commuting subsets of the Dirac observable algebra. This has still to be addressed within the proposal of \cite{DonnFreid}, as was also remarked in \cite{DonnellyGiddings2016}.
Perhaps, the relevant suggestion is that this splitting of the observable algebra might not be only achieved by removing certain observables, but also allowing for (many) more observables via the introduction of  a new unphysical `boundary field'.  
This additional boundary structure might, in fact, be used to construct new local observables which otherwise would not be available \cite{DonnellyGiddings2016}.

\section{Discussion}

Recent work has shown that the notion of entanglement entropy in gauge systems is ambiguous. The deep underlying reason is that due to the non-local features of the observable algebra in gauge systems, a notion of subsystems needs to be defined first. The way this question is answered does not only affect the definition of entanglement entropy, but has much wider implications for our understanding of (quantum) systems with gauge symmetries \cite{Oeckl2003,Giddings2015, DonnFreid,Gomes2016}. In particular, defining subsystems in background independent theories, {\it e.g.} gravity, leads to various completely open issues. The methods developed here lead to a new proposal for lattice gauge theories, that is also applicable to $(2+1)$ dimensional gravity. The main feature of this proposal is to use defect excitations to localize regions, which in the case of $(2+1)$ gravity means that regions are specified operationally by their particle content.

Furthermore, we clarified the relation between the different approaches put forward so far to define entanglement entropy, notably the extended Hilbert space approach \cite{Donnelly2014}, and the CHR approach \cite{Casini2013}, which focuses on the observable algebra.  In particular, we showed that the extended Hilbert space approach can be generalized to match not only the `electric centre' choice of \cite{Casini2013} but also the `magnetic centre' choice \cite{Casini2013}  (and its non-Abelian generalization). In our view, the resulting notion for subsystems can, in both approaches, be fully characterized by a choice of boundary conditions. In the non-Abelian case, the extended Hilbert space approach relies on the introduction of extra frame information at the boundary, which we argued could also be added in a generalized CHR approach. We have also seen that the proposal made here requires only the introduction of a global frame, which is then transported with a locally flat connection along the boundary.  In contrast, the spin-network based method of \cite{Donnelly2014} necessarily introduces for each link cut by the boundary---and in the continuum limit to each point of the boundary---a local frame. 
Nevertheless, we  observed that---if we wanted to---we could extend our framework as well, by allowing arbitrarily many frames along the boundary, leading to additional contributions to the entanglement entropy.

We have also pointed out that the different choices of boundary conditions, described by the `electric' vs. `magnetic' centre, are related to a choice of vacuum state. These vacuum states are of a `topological' nature, {\it i.e.} they arise as vacua of topological field theories with no local degrees of freedom. 
The states can be used to define continuum Hilbert spaces, that then describe the states of the related topological field theory with defect excitations.  Thus, for states describing  BF theory with defects, we have to choose the (generalized) `magnetic centre' definition, in order to obtain an entanglement entropy which is ($i$) regularization ({\it i.e.} graph) independent and ($ii$) finite.

The vacua we discussed here, are of a squeezed nature, which means they are sharply peaked either on flat connection (for the BF representation \cite{DGflux,DGfluxC,DGfluxQ} ) or vanishing electric fluxes (for the AL representation \cite{Ashtekar1986,Ashtekar1991,Ashtekar1993}).   The relation to preferred boundary conditions arises for the following simple reason: for states sharply peaked on connection degrees of freedom, it is natural and appropriate to fix the connection degrees of freedom (or the curvature) at the boundary. Similarly, for states peaked on some value of the electric flux, the original extended Hilbert space procedure \cite{Donnelly2014} based on spin-networks is the most natural and appropriate one.  Possible generalizations include $q$-deformed BF theory vacua \cite{DGTQFT}, corresponding to  $(2+1)$ gravity with a cosmological constant and, in condensed matter, to string net models \cite{Levin2004}. Furthermore, we suspect that also vacua with non-vanishing background values---{\it e.g.} for the electric fluxes \cite{Koslow2007,Koslow2011,Sahlmann2010}---come with a preferred notion of entanglement entropy. 

Our methods employ techniques from topological field theories, which in their `extended' form can be defined on manifolds with boundaries  and corners. We believe that this direction can be further explored in order to learn how to define the notion of subsystems, in particular for background independent systems.

\acknowledgements

We thank Rob Myers for encouragement and BD thanks Shinsei Ryu for discussions. 
CD is supported by an NSERC grant awarded to BD.
This research was supported in part by Perimeter Institute for Theoretical Physics. Research at Perimeter Institute is supported by the Government of Canada through the Department of Innovation, Science and Economic Development Canada and by the Province of Ontario through the Ministry of Research, Innovation and Science.

\newpage
\appendix
\section{Drinfel'd double \label{app_drinfeld}}
Drinfel'd doubles are examples of quasi-triangular Hopf algebras. They were presented in details in {\it e.g.} \cite{Koornwinder1999,Koornwinder1996,Dijkgraaf1991,Koornwinder1998,drinfel1988} and numerous identities were also proven in \cite{DDR1}. As a vector space, the Drinfel'd double of a group $\mathcal G$ is isomorphic to
\be
	\mD(\mG) \simeq \mathbb{C}[\mG] \otimes \mathcal{F}(\mG)
\ee
where $\mathbb{C}[\mG]$ is the group ring and $\mathcal{F}(\mG)$ is the Abelian algebra of linear functions on $\mG$. Therefore, a natural basis for $\mD(\mG)$ is provided by $\{G \otimes \delta_H\}_{G,H \in \mG}$ where $\delta_H$ is the delta function peaked on $H$, {\it i.e.} $\delta_H({\scriptstyle \bullet}) = \delta(H,{\scriptstyle \bullet})$. Henceforth, the more symmetric notation $[G,H] \equiv G \otimes \delta_H$ is  used. We now present some of the features of this algebraic structure. 

As a Hopf algebra, the Drinfel'd double is equipped with a multiplication rule
\begin{align}
	\star \;\; : \;\; \mD(\mG) \otimes \mD(\mG) \;\;\;& \longrightarrow \; \mD(\mG)\\
	\big( [\wG,\wH],[G,H]\big) \;&\longmapsto \; [\wG,\wH]\star [G,H] = \delta(\wH,\wG H \wG^{-1})[\wG G,\wH]
\end{align}
whose identity element is given by $\mathbb{I}=\sum_{H \in \mG}[e,H]$, as well as a comultiplication map
\begin{align}
	\Delta \;\; : \;\; \mD(\mG) \;& \longrightarrow \; \mD(\mG) \otimes \mD(\mG)\\
	[G,H] \;&\longmapsto \; \Delta[G,H] = \sum_{X,Y \in \mG \atop XY = H}[G,X] \otimes [G,Y] \; .
\end{align}
The irreducible representations $\{\rho\}$ of $\mD(\mG)$ follows the general construction of induced representations. Let $C$ be a conjugacy class of $\mG$ and $c_i$ its elements such that $c_1$ is the representative of the class. We define $N_C$ to be the stabilizer of the group element $c_1$ and we denote $\{R\}$ its complete set of irreducible representations. The corresponding matrix elements for $K\in \mG$ read $D^R_{M'M}(K)$ where $M',M$ are the magnetic indices. The elements of the quotient group $Q_C \simeq \mG / N_C$ are denoted $q_i$ and they satisfy the relation $c_i = q_i c_1 q_i^{-1}$. It turns out that the irreducible representations of $\mD(\mG)$ are labeled by a conjugacy class $C$ and an irreducible representation $R$ of $N_C$: $\rho = (C,R)$. The vector space on which the representation $\rho$ acts is denoted $V_{\rho}$. Furthermore, the matrix elements of the element $[G,H]$ in the representation $(C,R)$ are given by
\be
	D^{C,R}_{i'M',iM}([G,H]) = \delta(H,c_i')\delta(c_i',Gc_iG^{-1})D^R_{M'M}(q_{i'}^{-1}Gq_i) \;.
\ee
Thereafter, we make use of the more compact notation $D^{\rho}_{I'I} \equiv D^{C,R}_{i'M',iM}$ such that $\rho \equiv C,R$, $I' \equiv i'M'$ and $I \equiv iM$. The irreducible representations form a complete set of orthogonal representations. In particular, the orthogonality is provided by
\be
	\frac{1}{|\mG|}\sum_{G,H \in \mG}D^{\rho_1}_{I_1'I_1}([G,H])\overline{D^{\rho_2}_{I_2'I_2}([G,H])}
	= \frac{\delta_{\rho_1,\rho_2}}{d_{\rho_1}}\delta_{I_1',I_2'}\delta_{I_1,I_2} \; .
\ee
where $d_{\rho} = d_{C,R} = d_R . |C|$ denotes the dimension of the representation $\rho$ (denoted $\dim \rho$ or $\dim \rho_{{\mathcal D}(\mG)}$ in the main text).
Thanks to the comultiplication, tensor product of representations can be constucted:
\be
	(D^{\rho_1}\otimes D^{\rho_2})(\Delta[G,H]) = \sum_{XY\in \mG \atop XY = H}(D^{\rho_1}\otimes D^{\rho_2})([G,X]\otimes [G,Y]) \;.
\ee
Such tensor product can then be decomposed onto irreducible representations according to the fusion rules $N^{\rho_3}_{\rho_1 \rho_2}$ {\it i.e.}
\be
	\rho_1 \otimes \rho_2 = \bigoplus_{\rho_3} N^{\rho_3}_{\rho_1 \rho_2}\, \rho_3 \;.
\ee
This implies the existence of a unitary map $\mathcal{C}^{\rho_1 \rho_2}: \bigoplus_{\rho_3 \in \rho_1 \otimes \rho_2}V_{\rho_3}\rightarrow V_{\rho_1} \otimes V_{\rho_2}$ which satisfies
\be
	D^{\rho_1}_{I_1'I_1}\otimes D^{\rho_2}_{I_2'I_2}(\Delta[G,H]) = \sum_{\rho_3}\sum_{I_3'I_3}
	\mathcal{C}^{\rho_1\rho_2\rho_3}_{I_1'I_2'I_3'}\, D^{\rho_3}_{I_3'I_3}([G,H])\,\overline{\mathcal{C}^{\rho_1\rho_2\rho_3}_{I_1I_2I_3}},
\ee
{ where it is understood that the symbols ${\cal C}^{\rho_1\rho_2\rho_3}_{I_1'I_2'I_3'}$ represent matrices $\mathcal{C}^{\rho_1 \rho_2}$ whose indices} are given by the composed labels $\rho_3I_3'$ and $I_1'I_2'$. By analogy with the group case, such maps are referred to as Clebsch-Gordan coefficients. From the unitarity of $\mathcal{C}^{\rho_1 \rho_2}$, it follows the orthogonality relation
\be
	\sum_{I_1,I_2}\mathcal{C}^{\rho_1 \rho_2 \rho}_{I_1 I_2 I}\cdot \overline{\mathcal{C}^{\rho_1 \rho_2 \rho'}_{I_1 I_2 I'}} = \delta_{\rho,\rho'}\delta_{I,I'},
	\label{CGortho}
\ee
as well as the completeness relation
\be
	\sum_{\rho}\sum_I \mathcal{C}^{\rho_1 \rho_2 \rho}_{I_1' I_2' I}\cdot \overline{\mathcal{C}^{\rho_1 \rho_2 \rho}_{I_1 I_2 I}} = \delta_{I_1',I_1}\delta_{I_2',I_2} \;.
	\label{CGcomplete}
\ee

\section{Alternative fusion basis states \label{app_alt}}

In this appendix, we construct an alternative fusion basis for states defined on the four-punctured sphere. In \ref{subsub_fusion}, the following basis states were introduced
\begin{align}
	\psi_{\mathfrak{f}}^{\S_4}[\{\rho_i\}_{i=1}^5,\{I_k'\}_{k=1}^3,I_4]
 	&= \begin{array}{c}\includegraphics[scale =1]{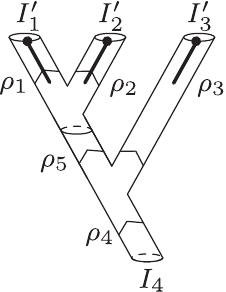}\end{array}  \\ 
	&= \sum_{\{I\}}\psi_{\mathfrak{f}}^{\S_2}[\rho_1,I_1'I_1]\psi_{\mathfrak{f}}^{\S_2}[\rho_2,I_2'I_2]\psi_{\mathfrak{f}}^{\S_2}[\rho_3,I_3'I_3]
	\; \mathcal{C}^{\rho_1 \rho_2 \rho_5}_{I_1 I_2 I_5}\; \mathcal{C}^{\rho_5 \rho_3 \rho_4}_{I_5 I_3 I_4}.
\end{align}
 These states form an orthonormal basis of the Hilbert space $\mathcal{H}_{p=4}$. An alternative basis is
\begin{align}
	\widehat{\psi}_{\mathfrak{f}}^{\S_4}[\{\rho_i\}_{i=1}^5,\{I_k'\}_{k=1}^2,\{I_k\}_{k=1}^2]
 	&= \mathcal{N}\begin{array}{c}\includegraphics[scale =1]{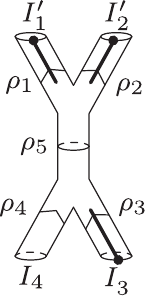}\end{array}  \\ 
	&= \mathcal{N}\sum_{\{I\}}\psi_{\mathfrak{f}}^{\S_2}[\rho_1,I_1'I_1]\psi_{\mathfrak{f}}^{\S_2}[\rho_2,I_2'I_2]\psi_{\mathfrak{f}}^{\S_2}[\rho_3,I_3'I_3]
	\; \mathcal{C}^{\rho_1 \rho_2 \rho_5}_{I_1 I_2 I_5}\; \overline{\mathcal{C}^{\rho_4 \rho_3 \rho_5}_{I_4 I_3' I_5}}\,,
\end{align}
where the factor $\mathcal{N}$ remains to be determined. To do so we ask the states $\widehat{\psi}_{\mathfrak{f}}^{\S_4}$ to be orthonormal. { The orthonormality is defined with respect to the inner product \eqref{innprod1} where the integral is over the holonomy variables implicitly represented by the bold edges. It follows from the orthonormality of the states $\psi_{\mathfrak{f}}^{\S_2}$ and the relation \eqref{CGortho} that the normalization factor must satisfy the identity}
\begin{align}
	\begin{array}{c}\includegraphics[scale =1]{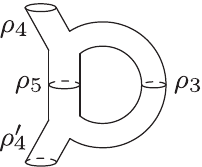}\end{array} = \mathcal{N}^{-2}
	\begin{array}{c}\includegraphics[scale =1]{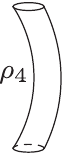}\end{array}\delta_{\rho_4,\rho_4'}
\end{align}
which after contracting both side with the identity $\delta_{I_4I_4'}$ gives
\begin{align}
	\mathcal{N}^{-2} = 
	\frac{\begin{array}{c}\includegraphics[scale =1]{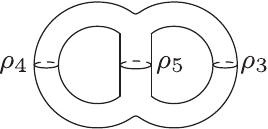}\end{array}}{\begin{array}{c}\includegraphics[scale =1]{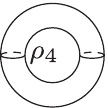}\end{array}}
	= \frac{d_{\rho_5}}{d_{\rho_4}}
\end{align}
where we used the unitarity (\ref{CGortho}) of the Clebsch-Gordan coefficients. Explicitly, this can be rewritten
\begin{align}
	\sum_{I_3,I_5}\mathcal{C}^{\rho_4 \rho_3 \rho_5}_{I_4 I_3 I_5}\; \overline{\mathcal{C}^{\rho_4' \rho_3 \rho_5}_{I_4' I_3 I_5}} = 
	\frac{d_{\rho_5}}{d_{\rho_4}}\delta_{\rho_4\rho_4'}\delta_{I_4I_4'}.
\end{align}
Putting everything together, the alternative fusion basis states read
\be
	\widehat{\psi}_{\mathfrak{f}}^{\S_4}[\{\rho_i\}_{i=1}^5,\{I_k'\}_{k=1}^2,\{I_k\}_{k=1}^2] = 
	 \sqrt{\frac{d_{\rho_4}}{d_{\rho_5}}}\sum_{\{I\}}\psi_{\mathfrak{f}}^{\S_2}[\rho_1,I_1'I_1]\psi_{\mathfrak{f}}^{\S_2}[\rho_2,I_2'I_2]\psi_{\mathfrak{f}}^{\S_2}[\rho_3,I_3I_3']
	\; \mathcal{C}^{\rho_1 \rho_2 \rho_5}_{I_1 I_2 I_5}\; \overline{\mathcal{C}^{\rho_4 \rho_3 \rho_5}_{I_4 I_3' I_5}} \; .
\ee
The same procedure can be easily employed in order to define the alternative fusion basis states $\widehat{\psi}_{\mathfrak{f}}^{\S_n}$ on the $n$-punctured sphere.
\section{States from ribbon operators \label{app_thecalcul}}
\begin{figure}[h]
	\includegraphics[scale =1]{fig/4punctminus-eps-converted-to.pdf} \q \q
	 \includegraphics[scale =1]{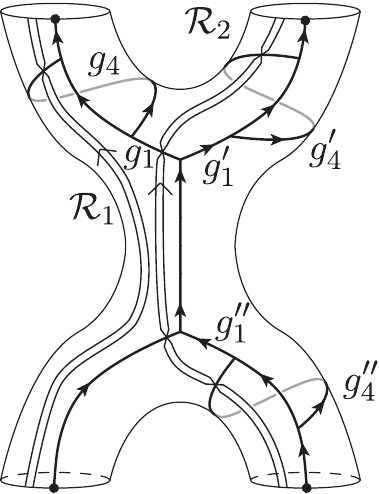} 
	\caption{The left panel depicts the situation where two non-intersecting ribbon operators go from a region $B$ to a region $A$ respectively defined as a set of two punctures on a four-punctured sphere. The right panel represents a topology equivalent to a four-punctured sphere. The solid lines represent the graph embedded on the surface while the double lines correspond to the ribbon operators acting on the links they cross. We can perform a gauge fixing by setting the group variables which are not labelled on the drawing to the identity.}
	\label{fig_4punctribbons}
\end{figure}

In this appendix, we study the states which are generated from the BF vacuum by non-intersecting ribbon operators going from a region $B$ to a region $A$. In particular, we are interested in the case of two ribbon operators $\mathcal{R}_1[\rho_1]$ and $\mathcal{R}_2[\rho_2]$ acting on the vacuum state defined on the four-punctured sphere as depicted in figure \ref{fig_4punctribbons}.

The starting point is the vacuum on the four-punctured sphere which is expressed in its gauge fixed form as
\be
	{\psi_0^{\mathbb{S}_4}}_{|\rm g.f.}\,=\,  \delta(g_4,e) \delta(g_4',e) \delta(g_4'',e)  \; .
\ee
We then apply two charge ribbon operators 
\begin{align} 
	&\big( \mathcal{R}_1[\rho_1,I_1'I_1]\mathcal{R}_2[\rho_2,I_2'I_2]\psi_0^{\S_4}\big) \\ \nn
	&\q = \frac{1}{|\mG|^2} \sum_{G_1,H_1 \atop G_2,H_2}\sqrt{d_{\rho_1}d_{\rho_2}}
	D^{\rho_1}_{I_1'I_1}([G_1,H_1])D^{\rho_2}_{I_2'I_2}([G_2,H_2])
	\big( \mathcal{R}_1[G_1,H_1]\mathcal{R}_2[G_2,H_2]\psi_0^{\S_4}\big)\; .
\end{align}
Now, the action of the (holonomy) ribbon operators on the vacuum state gives
\begin{align}
	&\big( \mathcal{R}_1[G_1,H_1]\mathcal{R}_2[G_2,H_2]\psi_0^{\S_4}\big)_{|\rm g.f.} = \delta(G_1,g_1)\delta(H_1,g_4)\delta(H_2,g_4')\delta(G_2,g_1'g_1'')\delta(H_2,G_2g_4''G_2^{-1}) \;,
\end{align}
which through the multiplication in ${\cal D}(\mG)$,  $[\wG,\wH]\star[G,H] = \delta(\wH,\wG H \wG^{-1})[\wG G,\wH]$, can be rewritten as
\begin{align}
	&\big( \mathcal{R}_1[G_1,H_1]\mathcal{R}_2[G_2,H_2]\psi_0^{\S_4}\big) = 
	\sum_{J_2^a\atop J_2^aJ_2^b = G_2} \big( \mathcal{R}_1[G_1,H_1]\mathcal{R}_2^a[J_2^a,H_2]\star \mathcal{R}_2^b[J_2^b,(J_2^a)^{-1}H_2J_2^a]\psi_0^{\S_4}\big)\,.
\end{align}
Here, the ribbon operator $\mathcal{R}_2^a$ ($\mathcal{R}_2^b$) goes along the link $l_1'$ ($l_1''$, respectively) and intersects the link $l_4'$ ($l_4''$, respectively). {Using the same factorization for the Drinfel'd double elements inside the representation matrices as for the ribbon operators, we can rewrite the action of the charge ribbons as }
\begin{align}
	&\big( \mathcal{R}_1[\rho_1,I_1'I_1]\mathcal{R}_2[\rho_2,I_2'I_2]\psi_0^{\S_4}\big) \\ \nn
	&\q =  \frac{1}{|\mG|^2} \sum_{{G_1,H_1 \atop G_2,H_2} \atop J_2^a}\sum_{I_2''}\sqrt{d_{\rho_1}d_{\rho_2}}
	D^{\rho_1}_{I_1'I_1}([G_1,H_1])D^{\rho_2}_{I_2'I_2''}([J_2^a,H_2])D^{\rho_2}_{I_2''I_2}([J_2^b,(J_2^a)^{-1}H_2J_2^a]) \delta(J_2^aJ_2^b,G_2)\\[-1.5em]
	& \q \q \q \q \q \q \q \q\times \big( \mathcal{R}_1[G_1,H_1]\mathcal{R}_2^a[J_2^a,H_2]\star \mathcal{R}_2^b[J_2^b,(J_2^a)^{-1}H_2J_2^a]\psi_0^{\S_4}\big) \; .
\end{align}
At this point it is useful to invoke the following identity
\be
	\sum_{I_2''}D^{\rho_2}_{I_2'I_2''}([J_2^a,H_2])D^{\rho_2}_{I_2''I_2}([J_2^b,\widetilde{H}_2]) = 
	\sum_{I_2''}D^{\rho_2}_{I_2'I_2''}([J_2^a,H_2])D^{\rho_2}_{I_2''I_2}([J_2^b,\widetilde{H}_2])\delta(\widetilde{H}_2,(J_2^a)^{-1}H_2J_2^a),
\ee
which shows that $ \widetilde{H}_2 =(J_2^a)^{-1}H_2J_2^a$ is automatically implemented by the contraction. This can be proven by inserting the resolution of the identity $\delta_{I'I} = \sum_{\wH \in \mG}\sum_{I''}D^{\rho_2}_{I'I''}([e,\wH])D^{\rho_2}_{I''I}([e,\wH])$ in the expression of the $\star$-multiplication in the Drinfel'd double representation $\rho_2$. Using this relation together with an obvious change of summation variables, we finally obtain
\begin{align}
	&\big( \mathcal{R}_1[\rho_1,I_1'I_1]\mathcal{R}_2[\rho_2,I_2'I_2]\psi_0^{\S_4}\big) \\ \nn
	&\q =  \frac{1}{\sqrt{d_{\rho_2}}}\frac{|\mG|}{|\mG|^3} \sum_{{G_1,H_1 \atop J_2^a,K_2^a}\atop J_2^b,K_2^b}\sum_{I_2''}\sqrt{d_{\rho_1}}d_{\rho_2}
	D^{\rho_1}_{I_1'I_1}([G_1,H_1])D^{\rho_2}_{I_2'I_2''}([J_2^a,K_2^a])D^{\rho_2}_{I_2''I_2}([J_2^b,K_2^b]) \\[-1.5em]
	& \q \q \q \q \q \q \q \q\times \big( \mathcal{R}_1[G_1,H_1]\mathcal{R}_2^a[J_2^a,K_2^a]\star \mathcal{R}_2^b[J_2^b,K_2^b]\psi_0^{\S_4}\big)\\
	& \q = \frac{|\mG|}{\sqrt{d_{\rho_2}}}\sum_{I_2''}
	\big( \mathcal{R}_1[\rho_1,I_1'I_1]\mathcal{R}_2^a[\rho_2,I_2'I_2''] \star \mathcal{R}_2^b[\rho_2,I_2''I_2]\psi_0^{\S_4}\big)
\end{align}
where in the last step use has been made of the inverse transform of \eqref{chargeribbon}.
 Notice also that at this point the $\star$ multiplication between ribbons in the last term is completely redundant.
 Moreover, one could have guessed this result directly from the $\star$ multiplication rules for charge ribbons. We preferred, however, a more explicit approach which makes clear the underlying path and graph structures, and the role of the generalized Fourier transforms as well.

Finally, we insert the resolution of the identity \eqref{CGcomplete}  provided by the Clebsch-Gordan coefficients between the path associated to the ribbon operators $\mathcal{R}_2^a$ and $\mathcal{R}_2^b$ and use the formula \eqref{ribbonstate} in order to obtain
\begin{align}
	&\big( \mathcal{R}_1[\rho_1,I_1'I_1]\mathcal{R}_2[\rho_2,I_2'I_2]\psi_0^{\S_4}\big) \\
	&\q = \frac{|\mG|}{\sqrt{d_{\rho_2}}}\sum_{\rho_3}\sum_{\{I\}}
	\psi_{\mathfrak{f}}^{\S_2}[\rho_1,I_1'I_1'']\psi_{\mathfrak{f}}^{\S_2}[\rho_2,I_2'I_2'']\psi_{\mathfrak{f}}^{\S_2}[\rho_2,I_2'''I_2]
	\; \mathcal{C}^{\rho_1 \rho_2 \rho_3}_{I_1'' I_2'' I_3}\; \overline{\mathcal{C}^{\rho_1 \rho_2 \rho_3}_{I_1 I_2''' I_3}} \\
	& \q = |\mG|\sum_{\rho_3}N^{\rho_3}_{\rho_1\rho_2}\sqrt{\frac{d_{\rho_3}}{d_{\rho_1}d_{\rho_2}}}
	\widehat{\psi}_{\mathfrak{f}}^{\S_4}[\rho_1,\rho_2,\rho_3,\rho_2,\rho_1;I_1',I_2',I_1,I_2] \;. \label{ribbonstates} 
\end{align}
where the state $\widehat{\psi}_{\mathfrak{f}}^{\S_4}$ was defined in appendix \ref{app_alt}. Note that the states \eqref{ribbonstates} are not not normalized and we therefore need to divide them by $|\mG|$.
\bibliography{Entanglement_final}
\end{document}